\documentclass[12pt]{article}
\usepackage{graphicx}
\usepackage{hyperref}
\usepackage{amsmath}
\usepackage{amscd,amssymb}
\usepackage[cmtip,arrow,%curve,
matrix]{xy}

\usepackage{curves}

\newcommand{\cE}{\mathcal{E}}
\newcommand{\cF}{\mathcal{F}}

\renewcommand{\P}{\mathbb{P}}

\DeclareMathOperator{\Td}{Todd}

\newcommand{\Z}{\mathbb{Z}}

\def\cR{{\mathcal R}}

\def\cN{{\mathcal N}}

\def\cC{{\mathcal C}}

\def\cH{{\mathcal H}}
\def\cP{{\mathcal P}}

\def\cE{{\mathcal E}}
\def\cF{{\mathcal F}}

\def\cO{{\mathcal O}}

\def\cZ{{\mathcal Z}}

\def\mC{{\mathfrak C}}
\def\mG{{\mathfrak G}}

\def\mS{{\mathfrak S}}

\def\mg{{\mathfrak g}}
\def\mh{{\mathfrak h}}

\def\ii{{\,{\rm i}\,}}
\def\dd{{\rm d}}

\newcommand{\Tr}[1]{\:{\rm Tr}\,#1}
\def\e{{\,\rm e}\,}
\newcommand{\IZ}{\mathbb{Z}}
\newcommand{\IC}{\mathbb{C}}
\newcommand{\IP}{\mathbb{P}}

\newcommand{\IR}{\mathbb{R}}

\newcommand{\IH}{\mathbb{H}}
\newcommand{\beq}{\begin{eqnarray}}
\newcommand{\eeq}{\end{eqnarray}}
\numberwithin{equation}{section}

\newcommand{\frS}{{\mathfrak{S}}}
\def\ch{{\rm ch}}
\def\Td{{\rm Td}}
\def\frsl{{\mathfrak{sl}}}

\begin{document}

\vspace{.1in}

\begin{center}

{\Large\bf Quantum Black Holes, Elliptic Genera \\[6pt] and Spectral Partition Functions}

\end{center}

\vspace{0.1in}

\begin{center}
{\large
A. A. Bytsenko $^{(a),}$\footnote{aabyts@gmail.com},
M. Chaichian $^{(b),}$\footnote{masud.chaichian@helsinki.fi},
R. J. Szabo $^{(c),}$\footnote{R.J.Szabo@hw.ac.uk},
A. Tureanu $^{(b),}$\footnote{anca.tureanu@helsinki.fi}
}
  
\vspace{5mm}
$^{(a)}$ 
{\it
Departamento de F\'{\i}sica, Universidade Estadual de
Londrina\\ Caixa Postal 6001,
Londrina-Paran\'a, Brazil}

\vspace{0.2cm}
$^{(b)}$
{\it
Department of Physics, University of Helsinki\\
P.O. Box 64, FI-00014 Helsinki, Finland}

\vspace{0.2cm}
$^{(c)}$
{\it
Department of Mathematics, Heriot-Watt University\\ Colin
Maclaurin Building, Riccarton, Edinburgh EH14 4AS, UK\\ Maxwell Institute
for Mathematical Sciences, Edinburgh, UK\\ The Tait Institute, Edinburgh, UK}

\vspace{3mm}

\end{center}

\vspace{0.1in}
%~\\
\begin{center}
{\bf Abstract}
\end{center}

We study M-theory and D-brane quantum partition
functions for microscopic black hole ensembles within the context of
the AdS/CFT correspondence in terms of highest weight 
representations of infinite-dimensional Lie algebras, elliptic genera, and
Hilbert schemes, and describe their relations to elliptic modular forms. 
The common feature in our examples lie in the
modular properties of the 
characters of certain representations
of the pertinent affine Lie algebras, and in the role of spectral 
functions of hyperbolic three-geometry associated with $q$-series in
the calculation of elliptic genera. We present new
calculations of supergravity elliptic genera on local Calabi-Yau
threefolds in terms of BPS invariants and spectral functions, and also
of equivariant D-brane elliptic genera on generic toric
singularities. We use these examples to
conjecture a link between
the black hole partition 
functions and elliptic cohomology.

\vfill

\begin{flushleft}                                                                        
August 2013 \hfill{EMPG--13--13}
                                                                                        
\end{flushleft}

\newpage

\setcounter{footnote}{0}

{\baselineskip=12pt
\tableofcontents}

\bigskip

\section{Introduction}

In this paper we discuss applications of affine Lie algebra 
representations, elliptic genera and their generalizations, and
Hilbert schemes to the computation of M-theory and D-brane quantum
partition functions for microscopic black hole ensembles, together with their
connections to elliptic modular forms and spectral functions in the
context of AdS/CFT holography. The 
main techniques we use involve the 
Grothendieck-Riemann-Roch theorem, characters of affine Lie algebras, and 
spectral functions of hyperbolic three-geometry. 
In particular, we make use of the characteristic classes of 
foliations of Lie group representations, which allow us to derive interesting new results and 
important old results in
superconformal field theory.

Let us begin by explaining the general scheme.
Suppose that $\mathfrak g$ is the Lie algebra of a Lie group $G$. Let us 
consider the pair $({H}, G)$ of Lie groups, where $H$ is a closed
subgroup of $G$ with normalizer subgroup $N_H\subset G$. Then the 
pair $({ H}, G)$ with the discrete quotient group $N_H/{ H}$ corresponds to the inclusion ${\mathfrak 
g}\hookrightarrow W_n$, where $W_n$ is the Lie algebra of formal vector 
fields in
$n = {\rm dim} \,G/H$ variables, while the homogeneous space $G/H$
possesses a canonical $\mathfrak g$-{structure} $\omega$
\footnote{Following~\cite{B-R}, by a $\mathfrak g$-{\it structure} on
  a smooth manifold $X$ we mean a smooth 
one-form $\omega$ on $X$ with values in $\mathfrak g$ which satisfies the 
Maurer-Cartan equation $\dd\omega = -\frac12\, [\omega, \omega]$,
i.e. for any pair of vector fields $\xi_1, \xi_2$ on $X$, one has $\dd\omega(\xi_1, 
\xi_2) = - [\omega(\xi_1), \omega(\xi_2)]$.}
(see \cite{Fuks} for details). Combining this $\mg$-structure with the 
inclusion $\mg\hookrightarrow W_n$, one obtains a $W_n$-structure on
the quotient space
$G/\mG$ for any discrete subgroup $\mG$ of the Lie group $G$; this is precisely 
the $W_n$-structure which corresponds to the $H$-equivariant foliation
of $G$ by left cosets of $\mG$~\cite{Fuks}.
The homomorphism
$
{{\rm char}_\omega :} \, H^\bullet (W_n)\rightarrow H^\bullet (G/\mG,
{\mathbb R})
$
associated with characteristic classes of $W_n$-structures 
decomposes into the composition of two homomorphisms
$
H^\bullet (W_n)\rightarrow H^\bullet ({\mathfrak g})$ and $
H^\bullet ({\mathfrak g}) \rightarrow H^\bullet (G/\mG, {\mathbb R})
$; the first homomorphism is independent of $\mG$ and is induced by the 
inclusion
${\mathfrak g}\hookrightarrow W_n$ (see e.g.~\cite{Bytsenko13}), 
while the second homomorphism is independent of $ H$ and corresponds to the
canonical homomorphism which determines
the characteristic classes of the canonical $\mathfrak g$-structure $\omega$ on $G/\mG$.
If the group $G$ is semisimple, then the Lie algebra $\mathfrak g$ is 
unitary and $G$ contains a discrete subgroup $\mG$ for which
$G/\mG$ is compact; for appropriate choice of $\mG$ the kernel of the homomorphism
$H^\bullet (W_n)\rightarrow H^\bullet (G/\mG, {\mathbb R})$
coincides with the kernel of the homomorphism $H^\bullet 
(W_n)\rightarrow H^\bullet ({\mathfrak g})$. 

In our applications we shall  
consider a compact hyperbolic three-manifold $G/\mG$ with $G = SL(2, {\mathbb 
C})$. 
By combining the characteristic class representatives of field theory elliptic
genera with the homomorphism
$
{{\rm char}_\omega}
$, one can compute the elliptic genera (and hence our quantum
partition functions) in terms of the 
spectral functions of hyperbolic three-geometry. All of these structures arise quite
naturally in string theory, and are particularly clear and tractable when supersymmetry is 
involved.

We shall now describe the specific contents of this paper, and their
interrelationships with the general sheme described above. In the
following we study examples of quantum black hole partition functions
and show that all of them, although different in their nature, can be
expressed in a generic way by means of spectral functions of
hyperbolic three-geometry. This enables us to establish concrete
modular properties of the partition functions even in instances that
are not covered by $AdS_3/CFT_2$ holography, and thus it naturally
explains the modularity in these cases. Along the
way a number of new calculations of elliptic genera are presented.

A central concept in this paper is that of the elliptic genus and its 
generalizations which we introduce in Sect. 2, together with some
pertinent examples. We analyze its
realisation as one-loop partition functions of superconformal 
sigma-models on symmetric product orbifolds, and survey various
features of the orbifold resolutions by Hilbert schemes, particularly the way
in which they yield natural geometric constructions of
representations of affine Lie algebras. 

In Sect. 3 we consider microscopic black hole partition functions in
M-theory on a suitable attractor geometry in the context of the
$AdS_3/CFT_2$ correspondence. We discuss its relationship to the minimal 
three-dimensional quantum gravity in a spacetime which is asymptotic to $AdS_3$;
the symmetry group of $AdS_3$ gravity (with appropriate boundary 
conditions) is generated by the Virasoro algebra, and the one-loop 
partition function is indeed the partition function of a conformal field 
theory in two dimensions. We then derive in detail the 
supergravity elliptic 
genus on
$AdS_3\times S^2\times X$ in terms of the Ruelle-type
spectral functions, with $X$ a Calabi-Yau threefold; we explain how to rewrite these M-theory black hole partition 
functions geometrically in terms of BPS invariants and 
derive a spectral function formulation for it. 

In Sect.~4 we analyze the black hole partition functions in the
reduction to 
Type IIA string theory, and the realisation therein of the black hole
components of chiral primary states of the minimal $\cN=4$
superconformal algebra. 
The chiral primary fields of an ${\mathcal N} = 2$ superconformal 
field theory form an algebra; the proof of this fact~\cite{Lerche,Warner} shares many 
common features
with the Hodge theory of K\"{a}hler manifolds, and here the connection
arises since the underlying D0-brane superconformal field theory is holographically dual to Type~IIA string theory on
$AdS_2\times S^2\times X$. We evaluate the generating 
function for chiral primary states in terms of the Ruelle-type
spectral functions. The microscopic black hole 
entropy is then evaluated from various points of view including the
one-loop partition function of a
generalised sigma-model, the Hirzebruch $\chi_y$-genus, and the
$(2,2)$ field theory elliptic genus. These results tie in to the
M-theory analysis of Sect.~3 through the relationship between the
equivariant genera of D-brane moduli spaces, which are given by
Hilbert schemes, and the
Gopakumar-Vafa expansion of topological string partition functions of
some local Calabi-Yau geometries in terms of BPS invariants. In particular, we present new calculations of equivariant D-brane elliptic genera
on generic toric singularities.

Finally, in Sect. 5 we briefly summarise our main conclusions from
this analysis. Two appendices at the end of the paper are devoted to a
brief definition and summary of the main features of the spectral functions 
of hyperbolic
three-geometry, including their analytic properties, and their
relations to the singularities of elliptic genera. From these facts, we conjecture 
an intimate connection between the black hole partition functions and
elliptic cohomology.

\section{Superconformal field theories, elliptic genera and symmetric 
products}

\subsection{Elliptic genera}

\label{EllGenera}

Elliptic genera are the  natural
topological invariants of manifolds which generalize the classical 
genera. They
appear, for example, when one considers the supersymmetric indices
of superconformal vertex algebras. For mathematicians, elliptic
genera are regarded as invariants for spaces in a generalized
cohomology theory extending K-theory, called elliptic cohomology. For 
physicists,
elliptic genera are the one-loop string partition functions which 
capture important
refined information about the spectrum of supersymmetric states of
the underlying superconformal field theory. Moreover, when they are applied
to instanton moduli spaces of supersymmetric gauge theories in four
dimensions, they also encompass as special cases the Vafa-Witten
partition function~\cite{Vafa} for $\cN = 4$ gauge theory and
Nekrasov's partition function \cite{Nekrasov} for $\cN = 2$
gauge theory (see e.g.~\cite{Cirio} for a collection of results). In 
this paper we will
exploit elliptic genera in calculating quantum black hole 
partition functions.

In physics, a two-variable elliptic genus can be
associated with any two-dimensional $\cN = (2,2)$ superconformal field 
theory~\cite{Eguchi89,Witten,Kawai}; it is given by
\begin{equation}
\Tr_{\cH}\, (-1)^F\, y^{J_0}\, q^{L_0-c/24}\, {\overline
q}^{{\overline L}_0-c/24}\ ,
\end{equation}
where $q$ and $y$ are complex parameters,
$L_0$ (resp. ${\overline L}_0$) is the Virasoro generator of left
(resp. right) movers, and $J_0$ is the $U(1)$
charge operator of left moving modes. The trace is taken over the
states in the Ramond-Ramond sector of the Hilbert space $\cH$ of the 
superconformal field theory, and $F= F_L+F_R$ with $F_L$ (resp. $F_R$)
the fermion number of left (resp. right) movers. The elliptic genus is
invariant under smooth variations of the parameters of the field theory and
is useful for enumerating BPS states.

For a superconformal sigma-model having as target space a smooth compact 
complex manifold
$X$ of dimension $d$, the genus one partition function is defined as the 
trace over the Ramond-Ramond sector of the time evolution operator
$q^{H}$ times $(-1)^F\, y^{F_L}$. If $\cE= \IC/(\IZ+\IZ\,\tau)$ is an elliptic curve
with complex structure modulus $\tau\in\IC^+$ and the line bundle parametrized by
$z\in {\rm Jac} (\cE)\cong \cE$, then we set $q := \e^{2\pi
\ii\tau}$ ($\ii := \sqrt{-1}\, $) and $y := \e^{2\pi \ii z}$. The field 
theory
elliptic genus is then defined as \cite{Dijkgraaf97}
\begin{equation}
\chi(X; q, y) = \Tr_{\cH(X)}\, (-1)^F\, y^{F_L} \,
q^{L_0-d/8}\, \overline{q}^{\overline{L}_0-d/8}\ ,
\label{char}
\end{equation}
where $\cH(X)$ is the Hilbert space of the ${\cal N}=2$ superconformal 
field theory with target space~$X$.

The formula (\ref{char}) has the form of a trace over an irreducible 
representation of the Virasoro
algebra containing a vacuum state $|0\rangle$ of a certain weight $k$, with
$L_{0}|0\rangle = -k\, |0\rangle$, along with its Virasoro descendants
$L_{-n_{1}}\cdots L_{-n_{l}} |0\rangle$. By the modular invariance of the
path integral, the elliptic genus (\ref{char}) is almost a modular form
under the $SL(2,\IZ)$ transformations of $\tau\in\IC^+$, and by the spectral 
flow it
is almost an elliptic function under lattice translations of $z\in\IC$ by
$\IZ+\IZ\,\tau$. Unitarity implies that the elliptic genus is a weak
Jacobi form such that the contribution to the trace from momentum
modes of the bosonised $U(1)$ current can be separated into
theta-functions which are the Jacobi forms of weight $1/2$ on
the subgroups of $SL(2,\IZ)$.

The field theory elliptic
genus (\ref{char}) may be alternatively defined in terms of
characteristic classes of $X$~\cite{Kawai,Dijkgraaf97,Borisov00}. For 
any complex
vector bundle $V$ over $X$, we can define the invertible formal sums
\begin{equation}
\mbox{$\bigwedge$}_q V = \bigoplus_{k=0}^d \, q^k\,
\mbox{$\bigwedge$}^{k}V \qquad \mbox{and} \qquad
\mS_qV = \bigoplus_{k=0}^{\infty}\, q^k\, \mS^kV \ ,
\end{equation}
where $\mbox{$\bigwedge$}^kV$ (resp. $\mS^kV$)
denotes the $k$-th exterior (resp. symmetric) product of
$V$. Introduce the invertible bundle
\begin{equation}
{\mathcal Ell}(V; q, y) = y^{-d/2}\, \bigotimes_{n= 1}^\infty\, \left(
\mbox{$\bigwedge$}_{-y\, q^{n-1}}V \otimes
\mbox{$\bigwedge$}_{-y^{-1}\, q^{n}}V^* \otimes \mS_{q^n}V
\otimes \mS_{q^n}V^* \right)
\end{equation}
regarded as a formal power series in $(q,y)$ with coefficients in the 
complex
K-theory group $K^0(X)$.
\footnote{Alternatively, we can regard
${\mathcal Ell}(V; q, y)$ as a sheaf with a well-defined holomorphic
Euler characteristic $\chi(X,{\mathcal Ell}(V; q, y))=\sum_{j\geq0}\,
(-1)^j\, \dim\, {\rm Ext}^j(\cO_X,{\mathcal Ell}(V; q, y))$.}
It is straightforward to see that the K-theory operation
${\mathcal Ell}$ is multiplicative, in the sense that for any two
complex vector bundles $V$ and $W$ on $X$ one has
\beq
{\mathcal Ell}(V \oplus W; q, y) & = & {\mathcal Ell}(V;q,y)
\otimes {\mathcal Ell}(W; q,y) \ , \nonumber \\[4pt] {\mathcal Ell}(V \ominus W; q, y) & = & {\mathcal Ell}(V;q,y)
\otimes {\mathcal Ell}(W; q,y)^{-1} \ .
\eeq
The formal power series
\begin{eqnarray}
\chi(X; q, y) := 
\chi\big(X\,,\, {\mathcal Ell}(TX; q, y) \big) = \sum_{j=0}^d\,
(-1)^j\,\dim H^j\big(X\,,\, {\mathcal Ell}(TX;{q,y}) \big)
\end{eqnarray}
is a holomorphic function on $\IC^+\times \IC$, called the {\em elliptic genus} of $X$. By the
Grothendieck-Riemann-Roch theorem, it can be computed in terms of
characteristic classes as
\begin{eqnarray}
\chi(X;q,y) 
&=& \int_{X}\, {\rm ch}\big({\mathcal Ell}(TX; q, y) \big)\,  {\rm
  Td} (X) \nonumber\\[4pt] &=& y^{-d/2}\, \int_{X}\ \prod_{j= 1}^d \,
x_j\ \prod_{n= 1}^\infty\, \left[\,
\frac{\big(1-y\, q^{n-1}\, \e^{-x_j}\big)\, \big(1-y^{-1}\, q^n\, \e^{x_j}\big)}
{\big(1- q^{n-1}\,\e^{-x_j}\big)\,\big(1-q^n\,\e^{x_j}\big)}\, \right]
\, ,
\label{RRH}
\end{eqnarray}
where $\{x_j\}_{j=1}^d$ are the Chern roots of the complex tangent bundle
$TX$. For a Calabi-Yau
manifold $X$ of dimension $d$, the elliptic genus (\ref{RRH})
is a weak Jacobi form of weight zero and index
$d/2$~\cite{Kawai,Borisov00}.

For $q=0$ the elliptic genus (\ref{char}) enumerates the BPS states with $L_0=\overline{L}_0=d/8$; one has $
{{\mathcal Ell}}(TX; 0, y) = y^{-d/2}\, \mbox{$\bigwedge$}_{-y}(TX)
$ and the elliptic genus reduces to the Hirzebruch $\chi_y$-genus~\cite{Li-Liu04}:
\beq
\chi_y(X)\ =\ y^{d/2}\, \chi(X;0,y)&=& \sum_{{j=0}}^d\, (-y)^j\, \chi
\big(X\,,\, \mbox{$\bigwedge$}^jTX\big) \nonumber \\[4pt] &=& \sum_{j= 0}^d \, (-y)^j \
\sum_{k= 0}^d \, (-1)^k \, \dim H^k\big(X\,,\, \Omega_X^j \big)\,,
\label{chiysum}\eeq
so that
\begin{eqnarray}
\!\!\!\!\!\!\!\!\!\!
\chi(X;q,y)\!\! \!& = \!\! \!& y^{-d/2} \chi_y(X) + q y^{-d/2}\, \sum_{j=0}^d
\left\{(-y)^{j+1} 
\chi \big(X\,,\,\mbox{$\bigwedge$}^j TX\otimes TX\big) \right.
\\ 
& \!+\!\! &
\left. \!\!q (-y)^{j-1} \chi \big(X,\mbox{$\bigwedge$}{}^j TX\otimes T^*X \big)
 \! + \! q (-y)^{j} \chi \big(X,
\mbox{$\bigwedge$}^j TX\otimes
(TX\otimes T^*X)\big)\right\}\!\!
+ \! \ldots \nonumber 
\end{eqnarray}
By the Grothendieck-Riemann-Roch theorem we have
\beq
\chi_y(X) = \sum_{j=0}^{d} \, (-y)^j \
\int_{X}\, \ch\big(\Omega_X^j\big)\,  \Td(X) = \int_{X} \ \prod_{j=1}^{d}\, \left[\frac{x_j\, \big(1-y\,
  \e^{-x_j}\big)}{\big(1-\e^{-x_j}\big)}\right] \ .
\label{chiycharclass}
\eeq
For a Calabi-Yau manifold $X$ of dimension $d$, the
$\chi_y$-genus transforms as
$$
\chi_y (X)=(-1)^{r-d}\, y^r\, \chi_{y^{-1}} (X)
$$ 
for some $r$. This relation
can be derived from the Serre duality 
\begin{equation*}
H^j\big(X\,,\,{\mbox{$\bigwedge$}}^sTX\big)\cong
H^{d-j}\big(X\,,\, {\mbox{$\bigwedge$}}^{r-s}TX \big) \ .
\end{equation*}
\medskip

{\bf Points.} \ 
Let $X$ be a point. Then
$V$ is a finite-dimensional complex vector space, which we assume to be $\mathbb Z$-graded. If ${\mathcal Ell}(V; q, y) = \bigoplus_{k,l}\, y^k\, q^l \, {\mathcal E}_{kl}(V)$, then the
bi-graded vector space ${\mathcal Ell}(V) := \bigoplus_{k,l}\,
{\mathcal E}_{kl}(V)$ has the structure of an $\cN=2$ superconformal
vertex algebra. 

\medskip

{\bf Hypersurfaces.} \ 
Let $X$ be a compact complex manifold of dimension $d$ and let $M \subset X$ be a
smooth hypersurface.
Denote by $[M]$ the line bundle on $X$ associated to the divisor $M$.
The adjunction formula states that $N_{M/X} \cong [M]|_M,$ where
$N_{M/X} = TX|_M/TM$ is the normal bundle of $M$ in $X$. The exact
sequence $0 \rightarrow TM \rightarrow TX|_M \rightarrow 
N_{M/X} \rightarrow 0$ of bundles on $M$
can then be rewritten as
\begin{eqnarray*}
0 \ \longrightarrow \ TM \ \longrightarrow \ TX\big|_M \ \longrightarrow
[M]\big|_M \ 
\longrightarrow \ 0 \ .
\nonumber
\end{eqnarray*}
In K-theory, $TM = (TX \ominus [M])|_M$ is the {virtual tangent bundle
of $M$ in $X$}, and in terms of Chern roots $\{x_j\}_{j=1}^d$ we have\footnote{
Note that
$c(M) = c(X)\, c([M])^{-1}|_M$. Let 
$
c(V) = \prod_j\, c(V_j)^{(-1)^j}
$
for a $\mathbb Z$-graded vector bundle $V = \bigoplus_j\, V_j$.
We have
$c(M) = c(TX \ominus [M])|_M$, and
$
\chi(M) = \int_M \, c(M)
= \int_{X} \, c(M) \, c_1(N_{M/X})
= \int_{X}\, c(TX \ominus [M])\, c_1([M])
$,
where $c_1([M])$ is the first Chern class of $[M]$.
}
\begin{eqnarray}
\chi(M;q,y) & = & y^{-(d-1)/2} \, \int_{X} 
\prod_{n = 1}^\infty\, \left[
\frac{\big(1-q^{n-1}\,\e^{-c_1([M])}\big)\, \big(1-q^n\,
  \e^{c_1([M])}\big)}{\big(1-y\, q^{n-1}\, \e^{-c_1([M])}\big)\,
  \big(1-y^{-1}\, q^n\, \e^{c_1([M])}\big)}\right]
\nonumber \\
& \times & 
\prod_{j=1}^{d} \, x_j \ \prod_{n = 1}^\infty\, \left[
\frac{\big(1-y\, q^{n-1}\,\e^{-x_j}\big)\, \big(1- y^{-1}\, q^n\, \e^{x_j}\big)}
{\big(1-q^{n-1}\,\e^{-x_j}\big)\, \big(1-q^n\,\e^{x_j}\big)}\right]
\, .
\label{Chern1}
\end{eqnarray}
For $q=0$ one has
$
{{\mathcal Ell}}([M]; 0, y) = y^{-1/2}\, (1 \ominus y \, [-M])
$ and according to~\cite{Ma-Zhou},
\begin{equation}
\chi_{y}(M)
= \chi\big(X\,,\, \mbox{$\bigwedge$}_{-y}TX\otimes (1 \ominus
y\, [-M])^{-1}\otimes (1 \ominus [-M])\big) \ .
\end{equation}
In terms of Chern roots, we finally obtain
\begin{equation}
\chi_{y}(M) = \int_{X} \, \frac{\big(1-\e^{-c_1([M])}\big)}{\big(1 - y\,
  \e^{-c_1([M])} \big)} \ 
\prod_{j=1}^{d}\, \left[\frac{x_j\, \big(1 - y\,
    \e^{-x_j}\big)}{\big(1-\e^{-x_j} \big)}\right] \,.
\label{F1}
\end{equation}

\medskip

{\bf Complete intersections.} \ 
Suppose now that $V\rightarrow X$ is a holomorphic vector bundle
on $X$ of rank $r$. Let $s: X \rightarrow V$ be a holomorphic section
transverse to the zero section such that the zero set
$Y:=s^{-1}(0)$ is a complex submanifold of $X$,
with $N_{Y/X} \cong V|_Y$. From the exact sequence
$
0 \rightarrow TY \rightarrow TX|_Y \rightarrow N_{Y/X}
\rightarrow 0
$
and the multiplicative property of ${\mathcal Ell}$ it follows that \cite{Ma-Zhou}
\begin{eqnarray*}
{{\mathcal Ell}}(TY;q,y) \otimes {{\mathcal Ell}}(V|_Y;q, y) =
{{\mathcal Ell}}(TX|_Y;q,y) = {{\mathcal Ell}}(TX;q,y)\big|_Y\ ,
\end{eqnarray*}
so that ${{\mathcal Ell}}(TY;q,y) = {{\mathcal Ell}}(TX;q,y) \otimes
{{\mathcal Ell}}(V;q;y)^{-1}|_Y$ and hence
\begin{eqnarray*}
\chi\big(Y\,,\, {{\mathcal Ell}}(TY;q,y) \big) = \chi\big(Y\,,\,
{{\mathcal Ell}} (TX \ominus V;q,y)|_Y \big)\ .
\end{eqnarray*}
By tensoring the Koszul complex \cite{Gri-Har}
\begin{equation*}
\!\!\!\!\!\!
0 \ \longrightarrow  \ \cO_{X}(\mbox{$\bigwedge$}^r V^* ) \
{\longrightarrow} \ 
\cO_{X}(\mbox{$\bigwedge$}^{r-1}V^*) \ 
{\longrightarrow} \ \cdots\ 
{\longrightarrow} \ \cO_{X}(\mbox{$\bigwedge$}^1V^*) \
{\longrightarrow} \ \cO_{X}
\longrightarrow \ \cO_Y \ \longrightarrow \ 0
\end{equation*}
with the bundle ${{\mathcal Ell}}(T{X} \ominus V;q,y)$, we get the exact sequence
\begin{eqnarray*}
0 & \longrightarrow & \cO_{X}\big({{\mathcal Ell}}(T{X} \ominus
V;q,y)\otimes \mbox{$\bigwedge$}^rV^* \big) \ 
{\longrightarrow}\ \cO_{X}\big({{\mathcal Ell}}(T{X}
\ominus V;q,y)\otimes \mbox{$\bigwedge$}^{r-1}V^* \big) \
{\longrightarrow} \ \cdots \\
& {\longrightarrow} & \cO_{X}\big({{\mathcal Ell}}(T{X}
\ominus V;q,y) \otimes \mbox{$\bigwedge$}^{1}V^*\big) \ 
{\longrightarrow} \ \cO_{X}\big({{\mathcal Ell}}(T{X}
\ominus V;q,y) \big) \\
& \longrightarrow & \cO_{X}\big({{\mathcal Ell}}(T{X} \ominus V;q,y)
\big) \big|_Y \ \longrightarrow \ 0 \ .
\end{eqnarray*}
Then by taking the Riemann-Roch character of this complex
one gets \cite{Ma-Zhou}
\begin{equation}
\chi(Y; q, y) = \chi\big(Y\,,\, {{\mathcal Ell}}(TX \ominus V;q,y)|_Y \big)
= \chi\big(X\,,\, {{\mathcal Ell}}(TX \ominus V; q, y) \otimes
\mbox{$\bigwedge$}_{-1} V^* \big) \ .
\end{equation}
Denoting by $\{v_k\}_{k=1}^r$ the Chern roots of $V$, we thus have
\begin{eqnarray}
\chi(Y; q, y)
& = &
y^{-\dim Y/2} \, \int_{X} 
\prod_{k=1}^r \ \prod_{n = 1}^\infty\, \left[
\frac{\big(1-q^{n-1}\,\e^{-v_k}\big)\,\big(1-q^n\,\e^{v_k}\big)}{\big(1-y\,
q^{n-1}\,\e^{-v_k}\big)\,\big(1-y^{-1}\, q^n\,\e^{v_k}\big)}\right] 
\nonumber \\
& \times &
\prod_{j=1}^{d}\, x_j \ \prod_{n = 1}^\infty \, \left[                            
\frac{\big(1-y\, q^{n-1}\,\e^{-x_j}\big)\,\big(1- y^{-1}\,
q^n\,\e^{x_j} \big)}
{\big(1-q^{n-1}\,\e^{-x_j}\big)\,\big(1-q^n\,\e^{x_j} \big)}\right]
\, .
\label{Chern2}
\end{eqnarray}
Note that the last products in the integrands of (\ref{Chern1}) and
(\ref{Chern2}) are the same, as both have been calculated in terms of
Chern roots of the tangent bundle $TX$; the first product in the integrand of
(\ref{Chern2}) is given in terms of Chern roots of the vector
bundle $V\to X$. Setting $q = 0$ we get back a similar result to
(\ref{F1}) with
\beq
\chi_{y}(Y)
&=&  \chi\big(X\,,\, \mbox{$\bigwedge$}_{-y}TX\otimes
(\mbox{$\bigwedge$}_{-y} V^*)^{-1}\otimes \mbox{$\bigwedge$}_{-1}{V^*}\big)
\nonumber \\[4pt]
&=& \int_{X} \ \prod_{k=1}^r \, \left[\frac{\big(1-\e^{-v_k}\big)}{\big(1-y\,
    \e^{-v_k} \big)}\right] \
 \prod_{j=1}^{d} \, \left[\frac{x_j\,
     \big(1-y\,\e^{-x_j}\big)}{\big(1-\e^{-x_j} \big)}\right] \ .
\label{F2}
\eeq

\subsection{Superconformal sigma-models on symmetric products}
\label{Sigma}

As shown in Ref. \cite{Dijkgraaf97}, the elliptic genus of an
${\mathcal N}=2$ superconformal sigma-model on a symmetric
product orbifold ${\mathfrak S}^nX$ equates to the partition function of a
second quantized string theory on a space $X$ with $S^1$-action.
\footnote{
See~\cite[Sect. 1.1]{Dijkgraaf97} for a string theory interpretation of superconformal sigma-models on ${\mathfrak S}^nX$.
}
In string compactifications on manifolds of the form $X\times S^1$,
one can consider the configuration of a D-string wound $n$ times
around $S^1$, and bound to a D$p$-brane where $p= d+1$. The quantum
mechanical degrees of freedom of this D-brane configuration are
naturally encoded in a two-dimensional sigma-model on the $n$-th
symmetric product of $X$, which describes the transverse fluctuations
of the D-string; this construction is originally due to \cite{Vafa95,Bershadsky95,Strominger1}.

Let us therefore consider a sigma-model on the $n$-th
symmetric product ${\mathfrak S}^nX$ of a K\"{a}hler manifold $X$ of
dimension $d$,
which is the orbispace 
$$
\mS^n X = \big[X^n/\mS_n\big] := \underbrace{X\times\cdots\times X}_n
\,\big/ \,{\mathfrak S}_n \ ,
$$
where $\mS_n$ is the symmetric group of order $n$ acting by permuting
factors. Objects of the category of the {\rm orbifold} stack $[X^n/{\mathfrak S}_n]$ 
are the $n$-tuples $(x_1,\ldots,x_n)$ of points in $X$; arrows are elements of the form $(x_1,\ldots,x_n; \sigma)$, where $\sigma \in \mS_n$. The arrow $(x_1,\ldots,x_n; \sigma)$ has source $(x_1,\ldots,x_n)$ and target $(x_{\sigma(1)},\ldots,x_{\sigma(n)})$. 
This category is a groupoid: the inverse of $(x_1,\ldots,x_n; \sigma)$ is $(x_{\sigma(1)},\ldots,x_{\sigma(n)}; \sigma^{-1})$. 
The genus one partition function depends on the boundary conditions imposed on the fermionic fields. For definiteness, following \cite{Dijkgraaf97} we choose the boundary conditions such that the partition function 
$\chi({\mathfrak S}^nX; q, y)$ coincides with the elliptic genus
defined as in (\ref{char}). 

\medskip

{\bf Generating functions.} \ 
The Hilbert space of an orbifold string theory can be decomposed into twisted
sectors ${\cH}_{\gamma}$ which are labelled by the
conjugacy classes $\{\gamma\} $ of the orbifold group $\mS_n$.
For a given twisted sector one keeps only those states that are invariant under the
centralizer subgroup $G_{\gamma}$ of the element $\gamma$;
let ${\cH}_{\gamma}^{G_{\gamma}}$ be an invariant subspace
associated with $G_{\gamma}$. One can parametrize the conjugacy
classes $[\gamma]$ by using a set of partitions $\{n_j\}_{j=1}^{s(\gamma)}$,
$\sum_{j=1}^{s(\gamma)} \, n_j=n$, where $n_j$ is the multiplicity of
the cyclic permutation $(j)$ of $j$ elements in
the decomposition of $\gamma$: 
$
[\gamma]=\prod_{j=1}^{s(\gamma)}\, (j)^{n_j}.
$
For this conjugacy class the centralizer subgroup of a permutation
$\gamma$ is \cite{Dijkgraaf97}
\begin{equation}
G_{\gamma}= \mS_{n_1}\times \prod_{j=2}^{s(\gamma)} \, 
\big(\mS_{n_j}\rtimes {\Z}_j^{n_j}\big) \ ,
\label{gamma}
\end{equation}
where each subfactor $\mS_{n_j}$ and ${\mathbb Z}_j$ permutes the $n_j$ cycles $(j)$ and acts within each cycle $(j)$
respectively. 
The total orbifold Hilbert space ${\cH}(\mS^n X)$ takes the form \cite{Dijkgraaf97}
\begin{equation}
{\cH}(\mS^n X)=
\bigoplus_{[\gamma]}\, {\cH}_{\gamma}^{G_{\gamma}}
= \bigoplus_{[\gamma]} \ \bigotimes_{j = 1}^{s(\gamma)} \,
\mS^{n_j}{\cH}_{(j)}^{\Z_{j}}\ ,
\end{equation}
where $\mS^n{\cH}:= ({\cH}^{\otimes n} )^{\mS_n}$ and we have decomposed each twisted sector ${\cH}_{\gamma}^{G_{\gamma}}$ into a product over the subfactors $(j)$ of $n_j$-fold
symmetric tensor products.

Let $\chi(-;q,y)$ be the generating function for any
Hilbert (sub)space of a superconformal sigma-model;
in~\cite{Landweber} it is
shown that the generating function for ${\cH}(\mS^n X)$ coincides with the elliptic
genus of $X$. If $\chi\big({\cH}_{(j)}^{\Z_{j}};q,y\big)$ admits the expansion
$\chi\big({\cH}_{(j)}^{\Z_{j}};q,y\big) = \sum_{n\geq0,l}\, \kappa(j\, n,l)\,
q^n \, y^l$, then one has the product formula~\cite{Kawai,Dijkgraaf97}
\beq
\sum_{n=0}^\infty\, p^n\, \chi(\mS^nX;q,y)= \prod_{m>0\,,\,n\geq0 \,,\,l}\,
\big(1-p^m\, q^n\, y^l\big)^{-\kappa(m\, n,l)} \ ,
\label{Ellproduct}\eeq
where $p:=\e^{2\pi\ii\rho}$ with $\rho$ the complexified K\"ahler form
of $\cE$. One can enhance the $SL(2,\IZ)\times SL(2,\IZ)$
transformations of
the parameters $(\rho, \tau)$ together with the elliptic
transformations of the
Wilson line modulus $z$ by combining them into a $2\times 2$ matrix
$$
\Xi = 
\begin{pmatrix}
\rho & z \\
z & \tau 
\end{pmatrix}
$$
 belonging to the
Siegel upper half-plane of genus two, with ${\rm Im}\, \rho >0,\, {\rm
  Im}\, \tau >0$, and ${\rm det}\,
{\rm Im}\, \Xi >0$.
The Narain duality
group $SO(3,2;{\mathbb Z})$ is isomorphic to the Siegel modular
group $Sp(4, {\mathbb Z})$ which acts on the matrix $\Xi$ by linear fractional transformations
$\Xi \mapsto (A\, \Xi +B)\, (C\,\Xi + D)^{-1}$. Then the elliptic genus is
almost an automorphic form for $O(3,2;\IZ)$.

For $q=0$ the elliptic genus reduces to the $\chi_y$-genus of ${\mathfrak
  S}^nX$ and by (\ref{chiysum}) the generating series (\ref{Ellproduct}) becomes
\beq
\sum_{n=0}^\infty\, q^n\, \chi_y(\mS^nX)= \prod_{j=0}^d \
\prod_{n=1}^\infty\,
\big(1-y^{n+j} \, q^n\, \big)^{(-1)^{j+1}\, \chi^{j}(X)} \ ,
\label{chiyproduct}\eeq
where $\chi^j(X):= \sum_{k=0}^d\, (-1)^k\,h^{j,k}(X)$ and $h^{j,k}(X)$
are the Hodge numbers of $X$.

For trivial line bundles $y =1$, the elliptic genus of the symmetric product ${\mathfrak
  S}^nX$ degenerates to the Witten index of the supersymmetric field
theory which is given by the Euler number $\chi({\mathfrak S}^nX; q,
1)= \chi({\mathfrak S}^nX)$~\cite{Hirzebruch2,Vafa}. This identifies
the elliptic genus with the 
character of the bosonic Fock module for the Heisenberg algebra, which
is almost a modular form of weight $-\chi(X)/2$,
through the identity
\begin{equation}
\sum_{n= 0}^\infty\, q^n \, \chi({\mathfrak S}^nX)= {\hat \eta}(\tau)^{-\chi(X)} =
\prod_{n = 1}^\infty \, \big(1-q^n\big)^{-\chi(X)} \ ,
\label{character}\end{equation}
where $\hat\eta(\tau)$ is the Euler function which is related to the Dedekind function by
$\eta(\tau):=q^{1/24}\, \hat\eta(\tau)$.
A similar formula, associated to the (equivariant) orbifold Euler characteristic of the symmetric product, can be derived using the $\mathfrak{S}_n$-equivariant K-theory of $X^n$ by means of the expression 
$
\chi({\mathfrak S}^nX) = {\rm rank}\, K^0({\mathfrak S}^nX) - {\rm rank}\,  K^1({\mathfrak S}^nX)
$.

\subsection{Hilbert schemes and representations of affine Lie algebras}\label{HH}

The moduli space
of zero-dimensional subschemes of length $n$ in a nonsingular
quasi-projective variety $X$ over $\mathbb C$ is called the {\it
  Hilbert scheme} of $n$ points on $X$, and is denoted by $X^{[n]}$;
it parametrizes D0-branes in terms of ideal sheaves supported at points of $X$.
A simple example of a zero-dimensional subscheme is a collection of
distinct points; the length is equal to the number of points. When
points collide, more complicated subschemes appear as these configurations appear in families:
when two points collide, there are infinitely many
nearby points, which form a pair consisting of a point $x\in X$ and a
one-dimensional subspace of the tangent space $T_xX$. This marks the
difference between $X^{[n]}$ and the $n$-th symmetric product
$\mS^nX= [X^n/\mS_n]$ in which the information about the one-dimensional subspace is absent.

When ${\rm dim}\,X = 1$, the Hilbert scheme $X^{[n]}$ is isomorphic
to $\mS^nX\cong X^n$ under the Hilbert-Chow morphism. For ${\rm dim}\,X = 2$, the Hilbert-Chow morphism $\pi : X^{[n]}\rightarrow \mS^nX$ gives a
resolution of singularities of the symmetric product
$\mS^nX$~\cite{Fogarty}; in particular, $X^{[n]}$ is a nonsingular
quasi-projective variety of dimension $2n$.
This is in contrast with Hilbert schemes for ${\rm dim}\,X
> 2$. For a projective variety $X$ the scheme $X^{[n]}$ is also projective, as
follows from Grothendieck's construction of Hilbert schemes. Moreover,
$X^{[n]}$ has a holomorphic symplectic form whenever $X$ has one~\cite{Fujiki,Beauville}.

Some interesting non-compact examples include $X = {\mathbb C}^2$ and
$X = T^{*}\Sigma$, where $\Sigma$ is a Riemann surface. For these complex
surfaces there exists a ${\mathbb C}^{*}$-action on $X$ which naturally induces an action on $X^{[n]}$ \cite{Nakajima99}.
On any smooth complex surface $X$ one can construct a representation of products of Heisenberg and Clifford algebras on the direct sum of homology groups of all
components $\bigoplus_{n\geq0}\, H_{\bullet}\big(X^{[n]},\IC\big)$; thus, the
generating function can be interpreted as a character of the
Heisenberg algebra. In particular, when $X$ is a K\"ahler manifold,
the space $\bigoplus_{n\geq0}\, H^\bullet\big(X^{[n]},\IC \big)$ forms
a representation of the Heisenberg algebra generated by geometrically
defined sigma-model
creation and annihilation operators $\alpha_n^a$, where $a$ labels a
basis of the cohomology $H^\bullet(X, \IC)$.

The relations between $\mS^nX$ and
$X^{[n]}$ bear many similarities to the relations between string
theory and field theory. For example, the Montonen-Olive S-duality
conjecture implies that the generating function for the Euler numbers
of moduli spaces of instantons in four dimensions should have modular
properties. For $X$ a K3 surface, the Euler numbers of the moduli
spaces of instantons are the same as those of Hilbert schemes of
points on $X$. Strictly speaking, one should consider moduli spaces
  of stable sheaves instead of moduli spaces of instantons which are
  usually non-compact, and this point of view will be taken in what follows. Then G\"{o}ttsche's formula (\ref{GFP}) below gives the desired answer.
Moreover, the homology groups of moduli spaces of instantons on an ALE
space, i.e. the minimal resolution of a simple Kleinian singularity,
form an integrable representation of a Kac-Moody algebra~\cite{Nakajima99}. The modular properties of the characters of the
representation (equivalently string partition functions) can be described
within the framework of conformal field theory. Heisenberg algebras and other
affine Lie algebras could perhaps be understood in this way within the framework of heterotic/Type~IIA duality.

\medskip

{\bf G\"{o}ttsche's formula.} \ 
When $X$ is an algebraic surface, the formula
(\ref{character}) also
computes the topological Euler characteristic of the Hilbert scheme
$X^{[n]}$~\cite{Gottsche}, while (\ref{Ellproduct}) computes the
elliptic genus of $X^{[n]}$~\cite{Borisov05}.
More generally, the generating function of the Poincar\'{e}
polynomials $P_t \big(X^{[n]} \big)=\sum_{j=0}^4\, t^j \, b_j\big(X^{[n]}\big)$ of $X^{[n]}$ is given by the
product formula
\begin{eqnarray}
\sum_{n=0}^{\infty}\, q^n\, P_t\big(X^{[n]}\big)= \prod_{j=0}^4 \ 
\prod_{n= 1}^\infty \, \big(1-(-t)^{2n+j-2} \, q^n\big)^{(-1)^{j+1}\, b_j(X)} \ ,
\label{GFP}\end{eqnarray}
where $b_j(X)=\dim H_j(X,\IR)$ is the degree $j$ Betti number of $X$.

When $X$ is a smooth quasi-projective Calabi-Yau threefold, the
Hilbert scheme $X^{[n]}$ is generally not smooth and has a complicated
subscheme structure with branches of varying dimension. Nevertheless, in
this case it
is still possible to define \emph{virtual} cycles and a virtual Euler
characteristic $\chi^{\rm vir}\big(X^{[n]}\big)$, which computes the
degree zero Donaldson-Thomas invariants of $X$, enumerating D0-branes parametrized by ideal
sheaves supported at points of $X$. Then there is an
analog of
G\"ottsche's formula which reads as~\cite{BehrFant}
\beq
\sum_{n=0}^\infty\, q^n\, \chi^{\rm vir}\big(X^{[n]}\big) =
M(\tau)^{\chi(X)} = \prod_{n=1}^\infty\, \big(1-q^n\big)^{-n\,
  \chi(X)} \ ,
\label{GFP3}\eeq
where $M(\tau)$ is the MacMahon function which can be represented as a
certain vacuum correlation function of vertex operators for the Heisenberg algebra. More generally, the generating function for the virtual version of the
Poincar\'e polynomials $P_t^{\rm vir}\big(X^{[n]}\big)$ of the Hilbert
scheme $X^{[n]}$ is the analog of (\ref{GFP}) given
by~\cite{BBSz}:
\beq
\sum_{n=0}^{\infty}\, q^n\, P_t^{\rm vir}\big(X^{[n]}\big)= \prod_{j=0}^6 \ 
\prod_{n= 1}^\infty \ \prod_{m=0}^{n-1} \, \big(1-(-t)^{2m-n+j-2} \,
q^n\big)^{(-1)^{j+1}\, b_j(X)} \ .
\eeq

\section{M-theory partition functions and (4,0) elliptic genera\label{se:40genera}}

\subsection{Black hole partition functions in the M-theory frame\label{Mframe}}  

If $X$ is a nonsingular compact Calabi-Yau threefold, then four-dimensional BPS black holes in M-theory compactifications on $X\times
S^1$ can be represented
microscopically by wrapping an M5-brane with fluxes and Kaluza-Klein momentum
on $M\times S^1$, where $M$ is a divisor in $X$. The dimensionally
reduced M5-brane worldvolume theory is dual to the Ramond sector of a
$(4,0)$ superconformal field theory~\cite{Maldacena}; the Neveu-Schwarz sector of this conformal
field theory is holographically dual to 11-dimensional supergravity (and the
five-dimensional black ring~\cite{Bena}) on the attractor geometry
$AdS_3\times S^2\times X$. The black hole
microstates are represented by the supersymmetric ground states of
this $(4,0)$ $CFT_2$ which are counted by an appropriately defined
elliptic genus describing the BPS gravitational configurations
inside $AdS_3$.

The black holes in question can be viewed as 
excitations of black strings, whose near-horizon geometry is a three-dimensional
anti-de~Sitter space $AdS_3$. The Euclidean thermal $AdS_3$ has a
conformal elliptic curve $\cE$ as its boundary whose complex structure modulus is
$\tau$. The partition function of the dual boundary conformal field
theory $CFT_2$ depends on $\tau$. The black hole partition functions
are then expected to possess a modular symmetry which is 
identified with the mapping class group $SL(2,\IZ)$ of the boundary
elliptic curve; after a choice of line
bundle on $\cE$, they are also expected to possess an elliptic symmetry which is identified with large gauge
transformations of the three-form field $H$ that leads to a spectral
flow symmetry of the partition function for the black string.

Consider the two-dimensional $(4,0)$ superconformal field theory on
the boundary $\cE$ of $AdS_3$ which describes M5-branes wrapping compact
four-cycles in $X$~\cite{Maldacena}; we consider $\cE$ to be a product of two circles
$S^1\times S_m^1$. We are interested in charge configurations
consisting of M5-M2 brane bound states which carry momentum $q_0$ along $S_m^1$,
with a single M5-brane wrapping $M\times S_m^1$, where $M$ is a very
ample divisor in $X$, and M2-branes wrapping holomorphic curves $\cC$ in $M$. The
divisor $M$ has a cohomology class $c_1([M]) \in H^2(X,\IZ)$, which can be
expanded as $c_1([M]) =p^a\, \Sigma_a$, where $\Sigma_a$ is a basis for
$H^2(X,\IZ)$.\footnote{In this paper all integer cohomology groups are
  understood modulo their torsion subgroups.} Via the pullback of the embedding $M\subset X$ this induces
a basis for $H^2(M,\IZ)$ which for simplicity we also denote by
$\Sigma_a$. Then the M2-brane charges $q_a$ are given by expanding $\cC=q_a\,
\cC^a$ where $\cC^a$ are dual two-cycles to $\Sigma_a$ in the lattice
$\Gamma:= H_2(M,\IZ)$. They
correspond to the numbers of left-moving excitations of total momentum
$q_0$ on the M5-brane. In the basis $\Sigma_a$ the triple intersection
form of $X$ is
\beq
D_{abc}= \frac16\, \int_X\, \Sigma_a\wedge\Sigma_b\wedge\Sigma_c \ , 
\label{D}\eeq
while the intersection matrix of $M$ is given by
\beq 
C_{ab}=-\int_M\, \Sigma_a\wedge \Sigma_b =-6D_{ab}\ , \qquad \mbox{with}
\quad D_{ab}:=D_{abc}\, p^c \ .
\eeq
We denote the inverse matrices to $C_{ab}$, $D_{ab}$ by $C^{ab}$,
$D^{ab}$. Let $b_2^\pm(M)$ denote the dimensions of the
spaces of selfdual/anti-selfdual two-forms on $M$. Then there are
$b_2^+(M)$ right-moving scalars on $S^1_m$ from dimensional reduction of
the self-dual three-form field strength $H$, and by the Riemann-Roch
formula one has
\beq 
b_2^+(M) = 2D +\mbox{$\frac16$}\, (c_2)_a\,
p^a
\eeq
where $D=D_{abc}\, p^a\, p^b\, p^c$ and
$
(c_2)_a=\int_X\, \Sigma_a\wedge c_2(X). 
$
The Euler characteristic of the divisor
$M$ is also computed by the Riemann-Roch theorem as
\beq 
\chi(M)= 2+b_2^+(M)+b_2^-(M) = \int_M\, c_2(M) = 6 D +(c_2)_a\, p^a = c_L  \ ,
\eeq
which coincides with the central charge $c_L$ of the conformal field theory,
i.e. the number of left-moving chiral bosons in the M-theory
frame.

The black hole partition function in the canonical ensemble is defined by
\beq \label{BHZX}
\cZ_X(p,\phi) = \sum_{q_0,q_a}\, d(q,p) \ \e^{-q_0\,
  \phi_0-q_a\, \phi^a}\,,
\eeq
where the quantum degeneracy $d(q,p)$ of black holes is the Witten index
\beq \label{qdegWitten}
d(q,p) = \Tr_{\cH_{q,p}(X)} \, (-1)^F
\eeq 
computed in the Ramond sector of the Hilbert space $\cH_{q,p}(X)$ of
the underlying $(4,0)$ superconformal
field theory with target space $X$. Here $\phi_0$ and $\phi^a$ are the respective chemical
potentials conjugate to the M-momentum and membrane charges. Being an index, it
receives only contributions from BPS states and is independent of the
background moduli. Once the black hole partition function is known,
the index of BPS states may be recovered from a suitable attractor contour
integral prescription:
\beq
d(q,p) = \int_{\sf C} \, \dd\phi_0\ \prod_{a} \, \dd\phi^a\ \e^{q_0\,
  \phi_0+q_a\, \phi^a} \ \cZ_X(p,\phi) \ ,
\label{Fouriercontour}\eeq
and the leading semi-classical approximation $\e^{S(q,p)}$ yields the
Bekenstein-Hawking-Wald entropy $S(q,p)$. For the partition functions
we compute in the following, the singularity and pole structures
required to evaluate such contour integrals are described in
Appendix~\ref{Pole}.

Later on we shall also consider black hole entropy counting
on certain classes of \emph{local} Calabi-Yau backgrounds $X$.
Such non-compact
Calabi-Yau manifolds can sometimes be thought of as decompactification
limits of compact
Calabi-Yau threefolds in neighbourhoods of collapsing cycles. In this
limit, gravity decouples so one needs to be careful about what is
meant by a ``black hole'' partition function. In the following, it will
always mean that the branes which microscopically realise the BPS black
hole wrap a non-compact cycle in the limit with nonzero entropy -- a BPS
``black hole'' is thus a BPS particle with large entropy. On non-compact Calabi-Yau manifolds $X$, a natural refinement of the
quantum black hole degeneracies (\ref{qdegWitten}) can be defined by replacing
the Witten index with a protected spin character that depends on an
extra parameter $y$ which gives a chemical potential conjugate to the angular momentum of states 
carrying the
$SU(2)$ R-symmetry that appears in the gravity decoupling
limit~\cite{Aganagic12}. The protected spin character is a supersymmetric index in M-theory, hence it also
receives contributions only from BPS black holes, and it is defined by
\beq 
d_y(q,p) = \Tr_{\cH_{q,p}(X)}\, (-1)^{2J_3}\, y^{J_3-J_0}
\ ,
\eeq
where $J_3$ is the generator of the $Spin(3)$ rotation group and $J_0$ is
the left-moving $U(1)$ R-charge operator of the four-dimensional BPS
states. This refined
index counts spinning M2-branes wrapping holomorphic curves $\cC$ in $X$,
and the refined black hole partition function is a two-variable
generalization of (\ref{BHZX}) defined by
\beq\label{BHZXy}
\cZ_X(p,\phi,y) = \sum_{q_0,q_a}\, d_y(q,p) \ \e^{-q_0\,
  \phi_0-q_a\, \phi^a} \ .
\eeq

\subsection{Quantum gravity and spectral partition functions on $AdS_3$} 

In the light of the $AdS_3/CFT_2$ correspondence, we expect a
duality between spectral zeta-functions of Euclidean $AdS_3$ and
elliptic modular forms. In particular, we consider
spectral functions of hyperbolic three-geometry whose arguments take
values on the elliptic curve $\cE$, which is the conformal boundary of
$AdS_3$, and are related to the congruence
subgroup of the modular group $SL(2,\IZ)$ which fixes one of the spin
structures on $\cE$; they are connected to various modular-like forms
(in particular Poincar\'e series) and thereby provide a spacetime perspective on the link between
quantum gravity partition functions and characters
of conformal field theory. The simplest example is provided by the
Fock module character $\hat\eta(\tau)$
from Sect.~\ref{Sigma} which can be expressed in terms of Selberg-type spectral
Ruelle functions (see Appendix~\ref{Spectral}) as 
\begin{equation}
\hat\eta(\tau)=\cR\big(s= 1- \ii\varrho(\tau)\big) \ ,
\label{characterspec}\end{equation}
where $\varrho(\tau)= {\rm Re}\,\tau/{\rm Im}\,\tau$. These spectral
functions carry information about the holographically dual
three-dimensional field theory, as their zeroes coincide with the scattering resonance poles of the
Laplace operator on the hyperbolic three-manifold.

In the following, we will describe the roles of modular forms and spectral functions in black
hole entropy counting. We will see that our
black hole partition functions, in the guise of elliptic genera that compute one-loop supergravity partition functions, can be
expressed in terms of spectral functions that provide spectral flow and a kind of
modular invariance. Of particular
relevance is the fact that the field theory elliptic
genus (\ref{RRH}) can be written in terms of the
spectral Ruelle function $\cR(s)$ as
\begin{eqnarray}
\chi(X;q,y) & =  & 
y^{-d/2}\, \int_{X} \ \prod_{j= 1}^d \, x_j
\left[\,\frac{\cR\big({\widehat s}= (2\pi\ii z-{{x}}_j)\, (1-\ii\varrho(\tau))\big)}{\cR\big({\widehat s}
= -{{x}}_j\, (1-\ii\varrho(\tau)) \big)}\right]
\nonumber \\
&  \times &
\left[\,\frac{\cR\big({\widehat s} = (2\pi \ii(\tau-z)
+{{x}}_j)\, (1-\ii\varrho(\tau)) \big)}
{\cR\big({\widehat s} = (2\pi \ii\tau+{{x}}_j)\, (1-\ii\varrho(\tau))
\big)}\, \right] \ ,
\label{HRR} 
\end{eqnarray}
where $q=\e^{2\pi\ii\tau}$, $y=\e^{2\pi\ii z}$ and
$\widehat{s}=2\pi\ii\tau\, s$. Proceeding in this way, we shall find
that our elliptic genera can be reproduced in terms of Selberg-type
spectral functions of the asymptotic $AdS_3$ geometry, and thereby inherit the modular properties of the
characters of highest-weight representations of the affine Lie
algebras which underlie the holographically dual conformal field theories.

The role of elliptic modular forms and spectral functions of hyperbolic
geometry can also be seen directly at the level of quantum corrections to
gravity on $AdS_3$. One-loop corrections to three-dimensional gravity
in locally anti-de Sitter spacetimes are qualitatively similar to
black hole quantum corrections. The simple geometric structure
of three-dimensional gravity allows exact computations,
since its Euclidean counterpart is locally isometric to a
constant curvature hyperbolic space $\IH^3$. 
Then
the quantum corrections to the gravity partition function on $AdS_3$ can be rewritten in terms of spectral
Ruelle functions as \cite{Bo-By,BBE,BytsenkoNew}
\begin{equation}
{\cZ}_{AdS_3}(\tau, \overline{\tau})  = \prod_{n
=2}^{\infty}\, \big|1-q^n\big|^{-2} =  \big| \cR\big(s=
2-2\ii\varrho(\tau)\big) \big|^{-2} \ .
\label{cZAdS3}\end{equation}

In the physics
literature it is usually assumed that the fundamental domain for
the action of a discrete isometry group $\mG^\gamma$ on
three-dimensional hyperbolic spacetimes has finite volume. In contrast, a three-dimensional black hole has a Euclidean
quotient representation $\IH^3/\mG^\gamma$ for appropriate $\mG^\gamma$,
where the fundamental domain has infinite hyperbolic volume; for
the non-rotating black hole one can choose $\mG^\gamma$ to be the
abelian group generated by a single hyperbolic element
\cite{Perry}. For the discrete groups of isometries of
three-dimensional hyperbolic space with fundamental domain of
infinite volume (e.g. for Kleinian groups, but excluding fundamental domains with
cusps), Selberg-type
functions and trace formulas are considered in \cite{Perry}. In an infinite-volume setting
the situation is complicated by the continuous spectrum of the
Laplacian, and the
absence of a canonical renormalization of the scattering operator for
the associated Schr\"odinger equation which is required
to render it trace-class. However, for a three-dimensional black
hole one can bypass much of the general theory and proceed
more directly, by defining a Selberg zeta-function attached to
$\IH^3/\mG^\gamma$ and establishing a trace formula which is a version
of the Poisson resummation formula for resonances. In fact, there is a special relation between the
spectrum and the {\it truncated} heat kernel of the Euclidean
black hole with the Patterson-Selberg spectral function
\cite{Bytsenko07}. 

\subsection{Supergravity boson-fermion correspondence\label{SUGRAbosferm}}

Let us consider in more detail the Neveu-Schwarz sector of the $(4,0)$ superconformal field theory
on the
$AdS_3$ boundary of the M-theory compactification on $AdS_3 \times S^2
\times X$. We shall compute the contribution to the elliptic genus of
the $(4,0)$ $CFT_2$ from supergravity states; they can be obtained, for
example, from the fluctuation spectrum of 11-dimensional supergravity
compactified on $AdS_3 \times S^2
\times X$. Later on we shall incorporate the contributions from
wrapped M2-branes and M5-branes with 
fluxes and momentum on cycles in $AdS_3\times S^2\times X$. We shall
make explicit reference to the
two left and right chiralities of the
$CFT_2$, with the convention that they correspond respectively to the holomorphic and
anti-holomorphic sectors. In addition to the underlying Virasoro
algebras, we pay close attention to the $U(1)$ and R-symmetry current
algebras.

To define the elliptic genus, we introduce chemical potentials
conjugate to the charges $q_a,\,\overline{q}{}_a$ as before; the R-charge is $q_0$. Up to a spectral flow, supergravity
states carry vanishing charges $q_a=0$; these charges are
instead carried by wrapped M2-branes, which we consider below. Hence the contribution to the
polar part of the elliptic genus from such supergravity states has the
schematic form $
\chi_{{\rm sugra}}(q) = \sum_n \, C(n)\, q^n$ with $n=q_0$ and
$q=\e^{-\phi_0}=: \e^{2\pi\ii\tau}$, and
in order to extract the quantum degeneracies $C(n)$ we have to
compute the genus contribution
$\chi_{{\rm sugra}}(q)$.

We work in the dilute gas expansion in which the contributions to the
elliptic genus are
dominated by multi-particle chiral primary states of massless
supergravity modes of M-theory compactified on $X$, i.e. supersymmetric ground states of BPS
hypermultiplets on $AdS_3\times S^2$. Their spectrum on $AdS_3\times
S^2$ organises into short representations of $SL(2,\IR)\times
SU(1,1|2)$. Consider the finite-dimensional subalgebra of the minimal
$\cN=4$ superconformal algebra in the Neveu-Schwarz sector spanned by
the operators
$\{L_0,L_{\pm\,1},T_0^i,G_{\pm\,1/2}^i,
  \overline{G}{}_{\pm\,1/2}^i\}_{i=1,2,3}$. Then a state
  $|\psi\rangle$ is said to be a chiral primary if
  $G_{-1/2}^2|\psi\rangle=\overline{G}{}^1_{-1/2} |\psi\rangle=0$ and
  $G^a_{n+1/2}|\psi\rangle=\overline{G}{}^a_{n+1/2}|\psi\rangle=0$, for
  $n\geq0$ and $a=1,2$. The subspace of chiral primary states is
  denoted $\cH^\prime(X)$.

The Neveu-Schwarz sector elliptic genus is related to that of the Ramond sector by spectral flow
\cite{Kraus}; the contribution from supergravity states can be written as
\begin{equation}
\chi^{\rm NS}_{\rm sugra}(q) = \Tr_{\cH^\prime(X)} \,
(-1)^{\overline{q}_0}\, q^{L_0} \ .
\end{equation} 
The elliptic genus receives contributions from
right-moving chiral primary states obeying $\overline{h}= \frac12\, \overline{q}_0$
\cite{Kraus}, where $\overline{h}$ is the eigenvalue of Virasoro
operator $\overline{L}_0$.

Consider the single particle spectrum, and suppose that it starts at a
conformal weight
$h_{\rm min} = \overline{h}_{\rm min}+s$ for some $s$.\footnote{ 
The appropriate complete spectrum of single particle primary states
can be found in e.g.~\cite{Kraus}. Note that
multiparticle chiral primaries can be obtained by taking arbitrary
tensor products of single particle chiral primary states.} For the bosonic contributions we compute
\begin{equation}
\prod_{l, p=0}^\infty \ \sum_{m=0}^\infty\,  q^{m\, (h_{\rm min}+l+p)}= 
\prod_{l, p =0}^\infty \, \big(1-q^{\overline{h}_{\rm min}+s+l+p}
\big)^{-1} \ ,
\end{equation} 
where $m$ labels the number of
particles, $p$ labels the actions of the operators $(L_{-1})^p$ to produce Virasoro
descendant states, and $l$ labels the shift in conformal dimension
$\overline{h} = \overline{h}_{\rm min}+l$. Setting
$n=l+p+1$ and using Eq. (\ref{RU1}) from Appendix~\ref{Spectral} for the boson and
fermion contributions, we then get
\begin{eqnarray}
\!\!\!\!
\chi_{\rm sugra}^{\rm NS,bos}(q) & = & \prod_{n=1}^\infty {\big(1-
q^{h_{\rm min}-1+n}\big)^{-n}}  \ = \ \prod_{n=1}^\infty\, \big[\cR\big(s=(n+h_{\rm
min}-1)\, (1-\ii\varrho(\tau))\big) \big]^{-1} \, , 
\nonumber \\[4pt]
\!\!\!
\chi_{\rm sugra}^{\rm NS,ferm}(q) & = & \prod_{n=1}^\infty \,
\big(1-
q^{h_{\rm min}-1+n}\big)^n \ = \ \ \prod_{n=1}^\infty\, 
\cR\big(s=(n+h_{\rm min}-1)\, (1-\ii\varrho(\tau))\big)\ . 
\label{fermion}
\end{eqnarray}

In the case of five-dimensional supergravity obtained by
compactifying M-theory on the Calabi-Yau threefold $X$, the five-dimensional massless
spectrum can be described in the $\cN=2$ superfield formalism in terms of the numbers
of vector multiplets $n_V = h^{1, 1}(X)-1$, hypermultiplets $n_H=
2(h^{2,1}(X)+1)$, and gravitino multiplets $n_S$, in addition to the
graviton multiplet; the Hodge numbers $h^{j,k}(X)$ in this context
are the numbers of generators of
degree $(j,k)$ \cite{Boer}. 
From Eqs. (\ref{characterspec}), (\ref{cZAdS3}) and (\ref{fermion}) we then find the supergravity elliptic genus to be
\begin{eqnarray}
\chi_{\rm sugra}^{\rm NS}(q) & = & \cR\big(s= 2(1-\ii\varrho(\tau))
\big) \ \big[\cR\big(s= 1-\ii \varrho(\tau)\big) \big]^{{\mathfrak
N}(n_V, n_H, n_S)} 
\nonumber \\ 
& \times & 
\prod_{n=1}^\infty\, 
\big[\cR\big(s= (n+1)\, (1-\ii\varrho(\tau))\big)\big]^{-\chi(X)}\ ,
\label{chisugraNS}
\end{eqnarray}
where the number ${\mathfrak N}(n_V, n_H, n_S)$ depends on the
multiplets and $\chi(X)=2(h^{1,1}(X) - h^{2,1}(X))$ is the Euler
number of the Calabi-Yau
manifold $X$.

\subsection{Supergravity elliptic genus on $AdS_3\times S^2\times X$\label{SUGRAelliptic}} 

We shall now compute the full supergravity elliptic genus from M-theory
following Ref. \cite{Gaiotto07}. In the Neveu-Schwarz sector it is given by 
\begin{equation}
\cZ_{\rm sugra}(\tau, \phi) = \Tr_{\overline{L}_0 = J^3_R} \, 
(-1)^F \, q^{L_0} \, \e^{-q_a\, \phi^a} \ .
\end{equation}
In this formula we have dropped a factor $q^{-c_L/24}$ which corresponds
to the ground state energy of $AdS_3$, and the $L_0$ contributions entering here
are due entirely to wrapped membranes. There are two kinds of
contributions: one from wrapped membranes and one from
supergravity modes. 
For M2-branes wrapping a holomorphic curve $\cC$ in $X$, the total
spin of highest weight states arising from a hypermultiplet in the representation
$(j_L,j_R)$ of the five-dimensional little group $SO(4)= SU(2)_L\times SU(2)_R$ is
$J^3=\frac12\, q_a\, p^a+m_R+m_L+\ell+\frac12$, where $-j_{L,R}\leq
m_{L,R}\leq j_{L,R}$ and $\ell\geq0$ is the orbital angular momentum
on $S^2$; the range of $j_L$ is determined by the genus of $\cC$, while
$j_R$ is related to the weight of the Lefschetz action on the
deformation moduli
space of $\cC$, and $q_a\, p^a$ is the contribution from the
quantization of the self-dual three-form field. 

Let us first consider the contribution of membranes wrapping an
isolated rational genus zero curve with fixed charges
$(q_a)\in\Gamma^*:=\Gamma\setminus\{0\}$ and degeneracy $N_{(q_a)}$. In this
case there is no internal $(j_L, j_R)$ angular momentum contribution,
and we have
\begin{eqnarray}
\cZ_{\rm sugra}^{(0, 0)}(\tau, \phi) = \prod_{p^a,n>0} \, \big[\big(1- q^{n+\frac12\, q_a\, p^a}\, \e^{-q_a\,
  \phi^a} \big)\, \big(1- q^{n+\frac{1}{2}\, q_a\, p^a}\, \e^{
  q_a\, \phi^a}\big) \big]^{n\, N_{(q_a)}} \ .
\nonumber \end{eqnarray}
For higher genus curves $\cC$ and general angular momenta $(j_L, j_R)$
we have
\begin{eqnarray}
\cZ_{\rm sugra}^{(j_L, j_R)}(\tau,\phi) & = & 
\!\!\!\!\! \prod_{\stackrel{\scriptstyle
p^a,n>0}{-j_L\leq m_L\leq j_L}} \, \big[\big(1- q^{n+\frac{1}{2}\, q_a\,
p^a+2m_L }\, \e^{-q_a\,\phi^a}\big)
\nonumber \\
& \times &
\!\!\!
\big(1- q^{n+\frac12\,q_a\, p^a+2m_L }\, \e^{q_a\, \phi^a} \big)\big]
^{(-1)^{2j_R+2j_L}\, n\, (2j_R+1)\, N_{(q_a) , j_L, j_R}}\,, 
\nonumber
\end{eqnarray}
where the quantum degeneracies $N_{(q_a) ,j_L,j_R}$ of the spinning
M2-branes are related to the
Gopakumar-Vafa BPS invariants of $X$, as we discuss in more detail below.
Finally, the net contribution from massless neutral supergravity modes
including singletons is given~\cite{Gaiotto07} by $\hat\eta(\tau)\,
M(\tau)^{\chi(X)}$, analogously to
(\ref{chisugraNS}). By the G\"ottsche product
formulas from Sect.~\ref{HH}, this shows that these modes are essentially parametrized by the punctual Hilbert schemes $X^{[n]}$ and are hence enumerated by the degree zero
Donaldson-Thomas invariants of $X$.

The Legendre transform of our original black hole partition function (\ref{BHZX}) is
computed by the elliptic genus of the $(4, 0)$ superconformal field
theory which is given as a
Ramond sector trace
\begin{equation}
\cZ_{X}(\tau, \phi)  =  \Tr_{\cH(X)}\, 
(-1)^F\, q^{L_0-{c_L}/{24}}\, {\overline
q}^{{\overline L}_0-{c_L}/{24}}\, \e^{-q_a\, \phi^a} \ .
\end{equation}
Collecting all contributions together, in the dilute gas expansion around ${\rm Im}\,\tau\rightarrow
\infty$ one gets $\cZ_X(\tau,\phi)= q^{-c_L/24}\, \cZ_{\rm
  sugra}(\tau,\phi)$ and hence
\begin{eqnarray}
\cZ_{X}(\tau, \phi) & = & 
q^{-c_L/24}\, \hat\eta(\tau)\, M(\tau)^{\chi(X)} \ 
\prod_{\stackrel{\scriptstyle
(q_a) \in \Gamma^*}{p^a,n>0}} \ \prod_{\stackrel{\scriptstyle 2j_{L,R}\geq0}{
-j_L\leq m_L\leq j_L}} \, \big[\big(1- q^{n+\frac{1}{2}\, q_a\,
p^a +2m_L }\, \e^{-q_a\,\phi^a}\big)
\nonumber \\
& \times &
\big(1- q^{n+\frac12\,q_a\, p^a+2m_L }\, \e^{q_a\, \phi^a} \big)\big]
^{(-1)^{2j_R+2j_L}\, n\, (2j_R+1)\, N_{(q_a) , j_L, j_R}} \ .
\label{Sugra}
\end{eqnarray}
Using Eq. (\ref{RU1}) from Appendix~\ref{Spectral} we find the spectral function representation
\begin{eqnarray}
\!\!\!\!\!\!\!\!\!\!
\cZ_{X}(\tau, \phi)\!\!\! & = & \!\!
q^{-c_L/24} \, \cR\big(s= 1-
\ii\varrho(\tau)\big) \ \prod_{n=1}^\infty \, \bigg[ \big[\cR
\big(s = n\, (1-\ii\varrho(\tau))\big)\big]^{-\chi(X)} 
\nonumber \\ 
\!\! & \times &
\!\!\!\!
\prod_{(q_a) \in \Gamma^* \,,\,p^a>0} \ \prod_{\stackrel{\scriptstyle 2j_{L,R}\geq0}{
-j_L\leq m_L\leq j_L}}\! \big[ \cR\big(s = (n +
q_a\,(\mbox{$\frac{p^a}2-\widehat{\phi}\,^a$})+ 2m_L )\, (1-\ii\varrho(\tau))
\big) 
\nonumber \\
\!\! & \times &
\!\!\! \cR\big(s = (n + q_a\,(\mbox{$\frac{p^a}2+\widehat{\phi}\,^a$}) +2m_L )\,
(1-\ii\varrho(\tau)) \big)\big]^{(-1)^{2j_R+2j_L}\, (2j_R+1)\,
N_{(q_a) ,j_L,j_R}} \bigg]\!,
\end{eqnarray}
where $\widehat{\phi}\,^a:= \phi^a/{2\pi\ii\tau}$.

\subsection{Spectral partition functions for BPS invariants}
\label{Symmetric}

We shall now explain how to rewrite our M-theory black hole partition
functions geometrically in terms of BPS invariants for some local Calabi-Yau
geometries $X$, and derive a spectral function formulation for
it. This can be done by using the Gopakumar-Vafa
conjecture~\cite{Gopakumar} which rewrites the topological string
partition function of $X$ as a sum over BPS configurations of
M2-branes wrapping holomorphic curves $\cC$ in $X$ and the M-circle. In contrast to the
Gromov-Witten invariants of $X$ which are in general rational numbers,
the generating series in M-theory of these
invariants in all non-zero degrees and all genera has a particular form
determined by certain integer invariants. There have been various
proposals for the proof of this conjecture (see
e.g.~\cite{Katz,Hosono,Li-Liu04}); for irreducible curve classes $\cC$
the
Gopakumar-Vafa invariants coincide with the Pandharipande-Thomas
stable pair invariants~\cite{PandThomas} which enumerate D2-branes
parametrized by
stable
sheaves supported on $\cC$.

Let $\cF_X^{\rm top}(\tau,Q)$ be the generating series of Gromov-Witten
invariants of a Calabi-Yau three-fold $X$. It counts the number of stable maps of {connected}
domain curves to $X$ in any given non-zero two-homology class; because
of the existence of automorphisms, one has to perform a weighted count
by dividing by the orders of the automorphism groups, and hence
Gromov-Witten invariants are in general rational numbers. The
Gopakumar-Vafa conjecture asserts particular integrality properties of
$\cF_X^{\rm top}(\tau,Q)$, and in particular the existence of integers
$n_{(q_a)}^g=n_{(q_a)}^g(X)$, which enumerate BPS states of wrapped membranes in $X$ such that
\begin{equation}
\cF^{\rm top}_X(\tau,Q) =  \sum_{(q_a)\in\Gamma^*} \ \sum_{g =
  0}^\infty \ \sum_{k= 1}^\infty \, \frac1{k} \, 
n_{(q_a)}^g \, \big(2\sin (\pi\, k\, \tau)\big)^{2g-2}\, \e^{-k\,q_a\, \phi^a} \ ,
\label{Gopakumar}
\end{equation}
where for a given curve class $\cC$ labelled by $(q_a)\in \Gamma^*= H_2(X,\IZ)\setminus\{0\}$ there are
only finitely many non-zero $n_{(q_a)}^g \in\IZ$ in genera $g\geq0$. Here $\phi^a:= \int_{\cC^a}\, \omega$ are the
K\"{a}hler parameters of $X$.
Let us regard $q$ as an element of $SU(2)$ represented by the diagonal matrix
${\rm diag}(q , q^{-1})$; then the sine functions in (\ref{Gopakumar})
can be interpreted as characters of $SU(2)$ representations and we
define integers $N_{(q_a)}^g$ such that
$$
\sum_{g=0}^\infty\, (-1)^g\, n_{(q_a)}^g\,
\big(2\sin(\pi\,\tau)\big)^{2g} = \sum_{g=0}^\infty\, N_{(q_a)}^g \ 
  \sum_{i=0}^g\, q^{g-2i} \ ,
$$
which is understood as a change of integral basis in the
representation ring of $SU(2)$.
Analogously to~\cite{Li-Liu04}, we then have
\begin{eqnarray}
\cF^{\rm top}_X(\tau,Q) & = &
\sum_{ (q_a)\in\Gamma^*} \ \sum_{g = 0}^\infty \ \sum_{k= 1}^\infty\,
\frac{(-1)^{g-1}\,N_{(q_a)}^g }{k \, \big(q^{{k}/{2}} - q^{-{k}/{2}}
\big)^2 } \, \e^{-k\, q_a\, \phi^a} \
\sum_{i=0}^{g}\, q^{k\,(g-2i)}
\label{G-V} 
\nonumber \\[4pt]
& = & \sum_{ (q_a)\in\Gamma^*} \ \sum_{g = 0}^\infty \ \sum_{k=
  1}^\infty\, \frac{(-1)^{g-1}\,N_{(q_a)}^g }{k} \, \e^{-k\, q_a\, \phi^a} \
\sum_{n= 0}^\infty\, (n+1)\, q^{k\, n} \ \sum_{i=0}^{g} \, q^{k\,(g-2i+1)}
\nonumber \\[4pt]
& = & \sum_{ (q_a)\in\Gamma^*} \ \sum_{g = 0}^\infty \, (-1)^g \,
N_{(q_a)}^g \ \sum_{i=0}^{g} \ \sum_{n= 0}^\infty \, (n+1)\,
\log \big(1- q^{g-2i+n+1} \, \e^{ -q_a\, \phi^a} \big).
\end{eqnarray}
In calculating the last line in Eq.~(\ref{G-V}) we can use the Ruelle function $\cR(s)$
from Appendix~\ref{Spectral} and the identity
$$
\sum_{n= 1}^\infty \, n\, {\rm log}\,\big(1-q^{n+\varepsilon} \big) = 
\sum_{n= 1}^\infty\, {\rm log} \, \cR
\big(s=(n+\varepsilon)\,(1-\ii\varrho(\tau)) \big)\,,
$$
for $\varepsilon\in\IC$. The generating series of {disconnected} Gromov-Witten invariants
is given by the topological string partition function
$\cZ^{\rm top}_X(\tau,\phi) = \exp (-\cF_X^{\rm top}(\tau,Q))$. Setting $j= g/2$ and
$N_{(q_a),j}:= N_{(q_a)}^g$, the Gopakumar-Vafa conjecture can be
reformulated as the infinite product formula (see also~\cite{Hollowood})
\beq 
\cZ_X^{\rm top}(\tau,\phi) &=& \prod_{(q_a)\in\Gamma^*\,,\, n>0} \
\prod_{\stackrel{\scriptstyle 2j\geq 0}{\scriptstyle -j\leq m\leq j}} \, 
\big[\cR\big(s=(n+2m+q_a\, \widehat{\phi}\,^a)\,
(1-\ii\varrho(\tau)) \big) \big]^{(-1)^{2j+1}\, N_{(q_a),j}} \nonumber
\\[4pt] &=& \prod_{(q_a)\in\Gamma^*\,,\, n>0} \
\prod_{\stackrel{\scriptstyle 2j\geq 0}{\scriptstyle -j\leq m\leq j}} \, \big(1-q^{n+2m}\, \e^{-q_a\,\phi^a} \big) ^{(-1)^{2j+1}\,n \, 
  N_{(q_a),j}} \ .
\label{eqn:GV}
\eeq
This formula reveals the geometric meaning of the membrane
contributions to the black hole
partition function (\ref{Sugra}).

\medskip

{\bf Resummation of the (4,0) elliptic genus.} \
We can turn this last calculation around to rewrite the expression
(\ref{Sugra}) for the black hole partition function in the form of
spectral functions, and rearrange it into a more geometric expansion in terms of BPS invariants
rather than the multiplicities
$N_{(q_a), j_L, j_R}$ of irreducible representations. Dropping the
ground state, neutral supergravity mode, and anti-membrane
contributions to the free energy
$\cF_X(\tau,\phi)=-\log\cZ_X(\tau,\phi)$ in Eq.~(\ref{Sugra}), we have
\begin{eqnarray}
\cF_X(\tau,\phi) & = & \sum_{\stackrel{\scriptstyle
(q_a)\in\Gamma^*}{\scriptstyle p^a,n>0}} \
\sum_{\stackrel{\scriptstyle 2j_{L,R}\geq 0}{\scriptstyle -j_L\leq
m_L\leq j_L}}\,  (-1)^{2j_L+2j_R}\, (2j_R+1)\, N_{(q_a), j_L, j_R}
\nonumber \\ 
& \times &  n\,
{\rm log} \,\big(1- q^{n + \frac12\,q_a\, p^a+2m_L}\,\e^{-q_a\,\phi^a}
\big)
\nonumber \\[4pt]
& = & 
- \sum_{\stackrel{\scriptstyle
(q_a)\in\Gamma^*}{\scriptstyle p^a,n,k>0}} \
\sum_{\stackrel{\scriptstyle 2j_{L,R}\geq 0}{\scriptstyle -j_L\leq
    m_L\leq j_L}}\,  (-1)^{2j_L+2j_R}\, (2j_R+1)\, N_{(q_a), j_L, j_R}
\nonumber \\ 
& \times & \frac nk\,q^{k\,(n +
 \frac12\,q_a\, p^a+2m_L)}\, \e^{-k\, q_a\, \phi^a}
\nonumber \\[4pt] 
& = &
- \sum_{\stackrel{\scriptstyle
(q_a)\in\Gamma^*}{\scriptstyle p^a,n,k>0}} \
\sum_{\stackrel{\scriptstyle 2j_{L,R}\geq 0}{\scriptstyle -j_L\leq
m_L\leq j_L}}\,  \frac{(-1)^{2j_L+2j_R}\, (2j_R+1)\, N_{(q_a),
j_L, j_R}\, n}{k\, \big(1-q^{k\, (\frac12\, q_a\, p^a+2m_L)}\,
\e^{-k\,q_a\, \phi^a} \big)} 
\nonumber \\ 
& \times &
\big(q^{k\, (n+\frac12\, q_a\, p^a+2m_L)}\, \e^{-k\, q_a\, \phi^a} -
q^{k\, (n+q_a\, p^a+4m_L)}\, \e^{-2k\, q_a\, \phi^a}  \big)
\nonumber \\[4pt]
& = &
\sum_{\stackrel{\scriptstyle
(q_a)\in\Gamma^*}{\scriptstyle p^a,n>0}} \
\sum_{\stackrel{\scriptstyle 2j_{L,R}\geq 0}{\scriptstyle -j_L\leq
m_L\leq j_L}}\,  (-1)^{2j_L+2j_R}\, (2j_R+1)\, N_{(q_a), j_L, j_R}
\nonumber \\ 
& \times & n\,
{\rm log} \, \left[\frac{\prod\limits_{l=n}^\infty\, \big(1-q^{l +
\frac12\,q_a\, p^a+2m_L} \,\e^{-q_a\,\phi^a}
\big)}{\prod\limits_{l = n}^\infty\, \big(1-q^{l+1 + \frac12\,q_a\, p^a+2m_L}\,\e^{-q_a\,\phi^a}
\big)}\right]
\nonumber \\[4pt]
& = &
\sum_{(q_a)\in\Gamma^*\,,\, p^a>0} \
\sum_{\stackrel{\scriptstyle 2j_{L,R}\geq 0}{\scriptstyle -j_L\leq
m_L\leq j_L}}\,  (-1)^{2j_L+2j_R}\, (2j_R+1)\, N_{(q_a), j_L, j_R}
\nonumber \\ 
& \times & 
\log\, \prod_{n=1}^\infty\, \big(1-q^{n +
\frac12\,q_a\, p^a+2m_L} \,\e^{-q_a\,\phi^a}
\big)^n
\nonumber \\[4pt]
& = &
\sum_{(q_a)\in\Gamma^*\,,\, p^a>0} \
\sum_{\stackrel{\scriptstyle 2j_{L,R}\geq 0}{\scriptstyle -j_L\leq
m_L\leq j_L}}\,  (-1)^{2j_L+2j_R}\, (2j_R+1)\, N_{(q_a), j_L, j_R}
\label{GF1} \nonumber \\ 
& \times & 
\log\, \prod_{n=1}^\infty\, \cR\big(s = (n +
q_a\,(\mbox{$\frac{p^a}2-\widehat{\phi}\,^a$})+ 2m_L )\, (1-\ii\varrho(\tau))
\big)\ .
\end{eqnarray}
The Gopakumar-Vafa invariants $n_r\big((q_a)\big)\in\IZ$ which count
M2-branes wrapping genus $r$ curves with charges $q_a$ are defined in
terms of the quantum degeneracies $N_{(q_a), j_L, j_R}$ via~\cite{Gopakumar,Gaiotto07}
\begin{equation}
\sum_{r=0}^\infty\, (-1)^r\, n_r\big((q_a)\big) \, \big(2\sin(\pi\,\tau)\big)^{2r} = \sum_{\stackrel{\scriptstyle 2j_{L,R}\geq 0}{\scriptstyle -j_L\leq
    m_L\leq j_L}}\, 
(-1)^{2j_L + 2j_R }\, (2j_R+1)\, N_{(q_a), j_L, j_R} \, q^{2m_L} \ .
\label{GF2}
\end{equation}
Hence BPS invariants of $X$ can be obtained by keeping only the
contributions from states with $j_L= 0$, and comparing
Eqs. (\ref{GF1}) and (\ref{GF2}) we obtain
\begin{eqnarray}
\cF_X^0(\tau,\phi) & = & \sin(2\pi\, \tau) \ \sum_{
(q_a)\in\Gamma^*\,,\, p^a>0} \ \sum_{r=1}^\infty \, (-1)^r\,
n_r\big((q_a)\big)\, \big(2\sin(\pi\,\tau)\big)^{2r} 
\nonumber \\ 
& \times & 
\sum_{n=1}^\infty\, \log\, \cR\big(s = (n +
q_a\,(\mbox{$\frac{p^a}2-\widehat{\phi}\,^a$}))\, (1-\ii\varrho(\tau))
\big) \ . 
\end{eqnarray}

\section{D-brane partition functions and (2,2) elliptic
  genera\label{se:22genera}}

\subsection{Black hole partition functions in the Type~IIA frame}

We shall now explain how to compute the contributions to the microscopic black hole
entropy in the setting of Sect.~\ref{se:40genera} by enumerating BPS bound states of D-branes inside the compact
Calabi-Yau threefold $X$. The same techniques are then employed for the computation of the supergravity elliptic genus in the
generalization of these considerations to BPS black holes in $\cN=2$
compactifications on more general Calabi-Yau manifolds $X$. One
particularly noteworthy feature of this formalism will be the natural appearance of
integrable highest weight representations of affine Lie algebras on the homology of
the D-brane moduli spaces, as anticipated from the $AdS_3/CFT_2$ correspondence. The characters of these modules compute the holomorphic parts of the torus
partition functions in the corresponding two-dimensional
superconformal field theory and are identified with black hole
partition functions as before.

Since the M5-brane wraps the circle $S_m^1$ and the M-momentum is
along $S_m^1$, we can take $S_m^1$ to be the M-circle and regard the
M5-M2 system from Sect.~\ref{se:40genera} as an 
M-theory lift of a
system of bound D4-D2-D0 states wrapping the divisor $M$ in Type~IIA
string theory compactified on the Calabi-Yau manifold $X$. These
supersymmetric ground
states can be regarded as excitations in the worldvolume theory on the
D4-brane. Below we will demonstrate how they may be regarded as multiparticle chiral primary
states in $AdS_3$. In the
large volume limit, this theory is the $U(1)$ Vafa-Witten
topologically twisted $\cN=4$ gauge theory~\cite{Vafa} on $M$. D0-branes are then
identified as instantons in this gauge theory which correspond to
massless supergravity modes, where the Kaluza-Klein
momentum on $S_m^1$ is the instanton number
\beq \nonumber
q_0=n=\frac1{8\pi^2}\, \int_M\, F\wedge F \ ,
\eeq
while the $U(1)$ fluxes on the D4-brane carried by D2-branes correspond to M2-branes
wrapping holomorphic curves on its worldvolume in
$X$ so that the membrane charges in the M-theory lift are magnetic monopole numbers
\beq \nonumber
q_a=C_{ab}\, u^b=\frac1{2\pi}\, \int_M\, F\wedge\Sigma_a \ .
\eeq
By self-duality one has
$q_a=0$ for all $a>b_2^+(M)$. In this setting, the Hilbert space of
BPS states $\cH_{q,p}(X)$ is the cohomology ring of the moduli space
${\mathcal M}_{n,u}(M)$ of $U(1)$ instantons on $M$ with second Chern class
$n$ and first Chern class $u$, and the degeneracy of bound states
\beq \nonumber
d(q,p) = \chi\big({\mathcal M}_{n,u}(M)\big)
\eeq
equals the Euler character of the instanton moduli space. The black
hole partition function (\ref{BHZX}) is then the generating function for these
Euler numbers given by
\beq \label{ZDtauz}
\cZ_M(\tau,z) = \sum_{n=0}^\infty \ \sum_{u\in \Gamma} \,
\chi\big({\mathcal M}_{n,u}(M)\big) \ q^n \ \prod_{a=1}^{b^+_2(M)}\, w_a^{u^a} \ ,
\eeq
where as before we set $q=\e^{-\phi_0}= \e^{2\pi\ii\tau}$ and we
define $w_a=\e^{-C_{ab}\, \phi^b}=:\e^{2\pi\ii z_a}
$. The smooth Gieseker
compactification of the instanton moduli space\footnote{Here we use
  the standard identification of D-branes with (torsion free) sheaves on
  large radius Calabi-Yau manifolds.} admits a factorization~\cite{Cirafici09}
\beq \label{MnuD}
{\mathcal M}_{n,u}(M) = M^{[n-n_u]} \times {\rm Pic}_{n_u}(M) \ ,
\eeq
where the punctual Hilbert scheme $M^{[n-n_u]}$ from Sect.~\ref{HH}
parametrizing D0-branes on $M$ is a smooth complex manifold of dimension $2(n-n_u)$, and ${\rm
  Pic}_{n_u}(M)$ is the Picard lattice which parametrizes D2-branes
corresponding to holomorphic line bundles on $M$ of first Chern class
$u$. The charge $n_u=\frac12\, u^a\, C_{ab}\, u^b+\frac12\, u^a\, c_1(M)_a$ is the curvature induced D0-brane charge on the
D4-brane. The sum over instanton numbers can be performed explicitly
by using G\"ottsche's formula from Sect. \ref{HH}, and in this way the black hole partition function 
(\ref{ZDtauz}) is computed to be~\cite{Fucito06,Griguolo06,Cirafici09}
\beq \label{BHZD}
\cZ_M(\tau,z) = \frac{\Theta_\Gamma
  (\tau,z)}{\hat\eta(\tau)^{\chi(M)}} = \Theta_\Gamma
  (\tau,z) \ \cR\big(s= 1- \ii\varrho(\tau)\big)^{-\chi(M)} \ ,
\eeq
where
\beq
\Theta_\Gamma (\tau,z) = \sum_{u\in \Gamma} \, q^{\frac12\, u^a\, C_{ab}\, u^b} \
 \prod_{a=1}^{b^+_2(M)}\, w_a^{u^a}
\eeq
is a Riemann theta-function on the magnetic M2/D2-brane charge lattice
$\Gamma =H^2(M,\IZ)$.

The refined black hole partition
function (\ref{BHZXy}) can be computed in the Type~IIA frame by considering the gauge theory on the full
five-dimensional D4-brane worldvolume $M\times S_m^1$. In this case,
instantons on the divisor $M$ become solitons on $M\times S_m^1$ whose worldlines
wrap around the circle $S_m^1$. Then the refinement parameter is
identified as $y=\e^{-\beta}=: \e^{2\pi\ii\rho}$, where $\beta$ is the radius of the Euclidean temporal circle $S_m^1$
that is also used to associate
a Hilbert space $\cH_{q,p}(X)$ with $M\times S_m^1$, which is
additionally graded by
angular momentum and R-charge. 
The refined black hole partition
function is now computed from an index in a supersymmetric quantum mechanics on the instanton moduli space
${\mathcal M}_{n,u}(M)$, which includes a product over all
Kaluza-Klein modes of the fluctuations around the D4-D2-D0 bound
states. In this case the protected spin character
\beq 
d_y(q,p) = \chi_y\big({\mathcal M}_{n,u}(M)\big)
\eeq
equals the Hirzebruch $\chi_y$-genus of the instanton moduli
space~\cite{Aganagic12}. Recall from Sect.~\ref{EllGenera} that it is
given by a sum of the form (\ref{chiysum}) over components of the BPS Hilbert
space, where the degrees $(j,k)$ are related to the spin and R-charge by
$(J_3,J_0) =\frac12\,
(j+k,j-k)$. In particular, for $y=0$ it computes the geometric genus
\beq 
\chi_0\big({\mathcal M}_{n,u}(M)\big) = \chi\big({\mathcal M}_{n,u}(M)\,,\,
\cO_{{\mathcal M}}\big)
\eeq
of ${\mathcal M} = {\mathcal M}_{n,u}(M)$, which gives the degeneracy of BPS states with equal spin and R-charge, while for $y=1$ we recover
the Witten index $\chi_1\big({\mathcal M}_{n,u}(M)\big) = \chi\big({\mathcal M}_{n,u}(M)\big)$. It can be computed in general in
terms of characteristic classes via Eq. (\ref{chiycharclass}) with
$X= {\mathcal M}_{n,u}(M)$ and $d=2(n-n_u)$.
The refined black hole partition function (\ref{BHZXy}) is then given by the
generating function
\beq 
\cZ_M(\tau,z,\rho) = \sum_{n=0}^\infty \ \sum_{u\in \Gamma} \,
  \chi_y\big({\mathcal M}_{n,u}(M)\big) \ q^n \ \prod_{a=1}^{b^+_2(M)}\,
  w_a^{u^a} \ .
\label{BHZDy}\eeq
From the perspective of the five-dimensional supersymmetric gauge
theory on the D4-brane, the counting parameter $q=\beta^2\, \Lambda^2$
is determined by the dynamical scale $\Lambda$ of the four-dimensional
gauge theory which arises in the dimensional reduction $\beta=0$. The
sum over instanton numbers can again be performed explicitly by using
results from Sect.~\ref{HH} and Eq.~(\ref{chiyproduct}) to get
\beq
\cZ_M(\tau,z,\rho) &=& \Theta_\Gamma(\tau,z) \ \prod_{j=0}^2 \
\prod_{n=1}^\infty\,
\big(1-y^{j+n}\, q^n\, \big)^{(-1)^{j+1}\, \chi^{j}(M)} \label{BHZDyproduct} \\[4pt]
&=& \Theta_\Gamma(\tau,z) \ \prod_{j=0}^2 \ \cR\big({\widehat s} =
2\pi \ii(\tau+j\, \rho)\, (1-\ii\varrho(\tau))-2\pi\ii\rho \big)^{(-1)^{j+1}\,
  \chi^{j}(M)} \ . \nonumber
\eeq

\subsection{Black hole components of chiral primary states} \label{Chiral}

We can alternatively consider the D4-D2-D0 states above from the
perspective of the D0-branes. This point of view can be used to study the black hole components of chiral primary fields of an ${\cN}=2$ superconformal
field theory in string theory compactified on the Calabi-Yau
threefold $X$; recall that chiral primary fields form an algebra~\cite{Lerche,Warner}.
D0-branes in the $AdS_{2}\times S^2\times X$ attractor geometry of an extremal
Calabi-Yau black hole with D4-brane charges $p^a$ are described by a
superconformal quantum mechanics~\cite{Gaiotto}. This superconformal theory contains a large degeneracy of chiral primary bound states. The degeneracy arises from D0-branes in the lowest Landau level which tile the
$ S^2\times X$ horizon, and such a multi-D0-brane conformal field
theory $CFT_1$ is holographically dual to Type~IIA string theory on
$AdS_2\times S^2\times X$. Of course, from the perspective of $AdS_2/CFT_1$ holography there is no \emph{a priori} reason
why the black hole degeneracy should exhibit any form of
modularity. On the other hand, modular properties are anticipated by
the (conjectural) S-duality of the D4-brane gauge
theory~\cite{Vafa}. Moreover, one expects that BPS states of D-branes
on Calabi-Yau manifolds have an algebraic structure akin to
generalised Kac-Moody algebras, generalising the geometric
construction of highest modules of affine Lie algebras that we
discussed in Sect.~\ref{HH}. 
Our rewriting below of the D-brane quantum partition functions in terms of
spectral Ruelle functions makes this modularity and algebraic
structure manifest.

We consider the case when a D2-brane wraps the
horizon $S^2$ and carries $N$ units of magnetic flux, corresponding to $N$ units of D0-brane charge. The D2-brane can be considered
as a point particle in $X$ with a two-form magnetic field
$c_1([M]) =p^a\, \Sigma_a$. The magnetic field divides the Calabi-Yau threefold $X$ into $D$
cells corresponding to the lowest Landau levels; in this setting the natural counting
function is the elliptic genus determined by Eq.~(\ref{Chern1}) with
\begin{eqnarray}
\chi(M;q,y) & = & y^{-1}\, \int_{X}\, 
\left[\frac{\cR\big({\widehat s}= -{{c_1}([M])}\,(1-\ii\varrho(\tau)) \big)}
{\cR\big({\widehat s}= (2\pi\ii z-{{c_1}([M])})\,
(1-\ii\varrho(\tau))\big)}\right] 
\nonumber \\ 
& \times &
\left[\frac{\cR\big({\widehat s}= (2\pi \ii\tau+{{c_1}([M])})\,
    (1-\ii\varrho(\tau)) \big)}
{\cR\big({\widehat s}= (2\pi \ii(\tau-z)+{{c_1}([M])})\,
  (1-\ii\varrho(\tau)) \big)}\right]
\nonumber \\
& \times &
\prod_{j=1}^{3} \, x_j\,
\left[\frac{\cR\big({\widehat s}= (2\pi\ii z-{{x}}_j)\,
    (1-\ii\varrho(\tau)) \big)}
{\cR\big({\widehat s}= -{{x}}_j\, (1-\ii\varrho(\tau)) \big)}\right]
\nonumber \\ 
& \times &
\left[\frac{\cR\big({\widehat s}= (2\pi \ii(\tau-z) +{{x}}_j)\, (1-\ii\varrho(\tau)) \big)}
{\cR\big({\widehat s}= (2\pi \ii\tau+{{x}}_j)\, (1-\ii\varrho(\tau))
\big)}\right]\, .
\end{eqnarray}
From our discussion of the modularity properties of the field theory
elliptic genus from Sect.~\ref{EllGenera}, this exhibits the D-brane
partition functions as weak Jacobi forms. However, here we shall instead proceed in a more direct way.

The chiral primary conditions can be written as
$
{{\nabla} \omega ={\nabla}^*\omega=0, \label{eq:chp}}
$
where $\omega$ is a $(j,k)$-form on $X$,
and $\nabla$ is a holomorphic covariant derivative with
connection that generates the magnetic field. Solutions of this equation are in one-to-one
correspondence with the elements of $H^k(X, \Omega_X^j\otimes
[M])$. By the Serre vanishing theorem, the cohomology group $H^k(X,
\Omega_X^j\otimes [M])$ vanishes for $k>0$ and
sufficiently large magnetic field
$c_1([M])$. 

The upshot is that we need compute the cohomology
$H^k(X, \Omega_X^j\otimes [M])$ for the moduli space
of a D2-brane on $X$, and the black hole entropy counting can be
reduced to a cohomology problem as before. In some cases one can
compute the dimension of $H^k(X, \Omega_X^j \otimes [M])$
using mirror symmetry,\footnote{
If $X$ is mirror to $Y$, then by homological mirror symmetry a D0-brane in $X$ is mirror to a three-torus $T^3$ in $Y$.
Moreover, the moduli space of a D3-brane wrapping such a $T^3$ in $Y$
is $X$. One can thus compute
$H^k(X, \Omega_X^j \otimes [M])$ and therefore the
black hole entropy using homological mirror symmetry.
} 
while 
$$
b_j:=\dim H^0\big(X\,,\, \Omega_X^j\otimes [M] \big)
$$ 
can be obtained by using
the Riemann-Roch formula to get (see also \cite{Maldacena,Gaiotto})
\begin{eqnarray}
b_0 & = & 
\int_X \, \Big(\,
\frac{c_1\big([M]\big)\wedge c_1\big([M]\big)\wedge c_1\big([M]\big)}{6}+\frac{c_2(X)\wedge
  c_1\big([M]\big)}{12} \, \Big) \ = \ {D} +
\mbox{$\frac1{12}$}\, (c_2)_a\, p^a \ = \ b_3 \ ,
\nonumber \\[4pt]
b_1 & = & \int_X\, \Big(\,
\frac{c_1\big([M]\big)\wedge c_1\big([M]\big)\wedge c_1\big([M]\big)}{2}-\frac{3c_2(X)\wedge
  c_1\big([M]\big)}{4}+\frac{c_3(X)}{2}\, \Big) \nonumber \\[4pt] &=& 3{D} -
\mbox{$\frac34$}\, (c_2)_a\, p^a-\mbox{$\frac12$}\, \chi(X) \ ,
\nonumber \\[4pt]
b_2 & = & \int_X\, \Big(\,
\frac{c_1\big([M]\big)\wedge c_1\big([M]\big)\wedge c_1\big([M]\big)}{2}-\frac{3c_2(X)\wedge c_1\big([M]\big)}{4}-\frac{c_3(X)}{2}
\, \Big) \nonumber \\[4pt] &=& 3{D} -\mbox{$\frac34$}\, (c_2)_a\, p^a
+\mbox{$\frac12$}\, \chi(X) \ .
\end{eqnarray}

We now have to count the multiparticle primaries, by first choosing a
basis of states. As in Sect.~\ref{Sigma}, we can partition the D0-branes into $k$
clusters, with the $l$-th cluster containing $n_l$ D0-branes where
$\sum_{l=1}^k\, n_l=N$.
Each cluster forms a wrapped D2-brane with $n_l$ units of magnetic
flux (we ignore the possibility of multiply wrapped D2-brane bound states). Then
each of the $k$ wrapped D2-branes can sit in one of the $b_0+b_1+b_2+b_3$ chiral primary states.
The counting of configurations is in fact in one-to-one
correspondence with the counting of states for a conformal field theory
with $b_0+b_2$ bosons and $b_1+b_3$ fermions of total momentum $n=n_l$.
The partition function of the appropriate conformal field theory can
be calculated by taking the trace over chiral primary states, with the result
\begin{equation}
\cZ_{\rm CFT}(\tau) =
\prod_{n = 1}^\infty \, \frac{\big(1+q^n\big)^{b_1+b_3}}
{\big(1-q^n\big)^{b_0+b_2}} = \frac{\big[\cR\big(s = 1- \ii\varrho(\tau) +
  \ii/(2\, {\rm Im}\,\tau)\big)
\big]^{b_1+b_3}}
{\big[\cR\big(s= 1-\ii\varrho(\tau) \big)\big]^{b_0+b_2}} \ .
\label{generating}
\end{equation}

\subsection{Microscopic black hole entropy}

We will now show that (\ref{BHZD}) is in perfect agreement with the microscopic
M5-brane computation of the black hole entropy in the two-dimensional
$(4,0)$ superconformal field theory derived by~\cite{Maldacena}; it also agrees with the macroscopic
supergravity result~\cite{Fucito06}. The macroscopic entropy of these black holes should coincide with the
asymptotic growth of the degeneracy by the Boltzmann relation
\beq 
S(q,p) = \log d(q,p)
\eeq
for $q,p\gg1$. On the other hand, the asymptotic growth of the microscopic degeneracy
evaluated via the Cardy formula gives the black hole entropy
\beq 
S(q,p) = 2\pi\, \sqrt{\mbox{$\frac16\, c_L\, \tilde q_0$}}  \ ,
\label{Cardy}
\eeq
where $\tilde q_0$ is the momentum remaining of the total M-momentum
$q_0$ to freely distribute in the M-theory frame; one can in fact improve the Cardy formula
(\ref{Cardy}) by including a degeneracy prefactor~\cite{BytsenkoNew}. 
In the present case one finds~\cite{Fucito06}
\beq 
S(q,p)= 2\pi\, \sqrt{\big(D +\mbox{$\frac16$}\,
(c_2)_a\, p^a\big)\, \big(q_0+\mbox{$\frac1{12}$}\, q_a\, D^{ab}\,
q_b\big) }
\eeq
for $q,p\gg1$, where the first term under the square root is $\frac16\, c_L= \frac16\,\chi(M)$ while
the second term is the number $\tilde q_0$ of D0-branes coming from the expansion of $\hat\eta(\tau)^{-\chi(M)}$
which gives the contributions from massless neutral supergravity
modes. 

\subsection{Representations of Heisenberg algebras}  

The expression (\ref{BHZD}) is also the partition function of a chiral two-dimensional conformal
field theory of $N=\chi(M)$ free bosons with $N_+=b_2^+(M)$ degrees of
freedom on the charge lattice $\Gamma =H^2(M,\IZ)$. We elucidate this
observation by calculating the black hole partition function from a
sigma-model based on a generalized chiral algebra~\cite{Bonelli12}. Consider the Heisenberg Lie algebra
$\widehat\mh(\Gamma)$ based on the lattice $\Gamma$ with the
intersection form $C_{ab}$. It has generators $\alpha_n^a$, $n\in\IZ$,
$a=1,\dots,N_+$ which satisfy the commutation relations
$
\big[\alpha_n^a,\alpha_m^b\big] = \delta_{n+m,0} \ C^{ab}.
$

Let $\cF(\Gamma)$ be the irreducible Fock space representation of
$\widehat\mh(\Gamma)$; by Sect.~\ref{HH} it can be identified geometrically with the cohomology ring of the
instanton moduli space $\bigoplus_{n,u}\,
H^\bullet\big({\mathcal M}_{n,u}(M),\IC \big)$. Define a sigma-model whose
target space is the degree two $L^2$-cohomology of $M$ modulo integral
elements, so that the momentum lattice is $\Gamma$, and whose
Hamiltonian is given by 
\beq 
L_0 = \frac12\, \alpha_0^a\, C_{ab}\, \alpha_0^b + \sum_{n=1}^\infty\,
\alpha_{-n}^a\, C_{ab}\, \alpha_n^b \ .
\eeq
Then the black hole partition function (\ref{BHZD}) can be
reproduced from the one-loop partition function of the sigma-model
\beq 
\chi_M(\tau,z) = \Tr_{\cF(\Gamma)}\, q^{L_0} \, \e^{-\zeta_a\, J_0^a} =
\hat\eta(\tau) \ \cZ_M(\tau,z)\,,
\eeq
where $\zeta_a=C_{ab}\, \phi^b=-2\pi\ii z_a$ and $J_0^a=\alpha_0^a$. In the M-theory frame, the extra free boson partition function
$\hat\eta(\tau)$ arises from the remaining part of the reduced
self-dual three-form field. The
generalized chiral algebra is thus associated with the sum of Heisenberg
algebras $\widehat\mh\oplus \widehat\mh(\Gamma)$.

\subsection{Toric Calabi-Yau black holes\label{Refined}}

The D-brane partition function (\ref{BHZD})
is derived in~\cite{Fucito06,Griguolo06,Cirafici09} for a large class of \emph{non-compact} toric manifolds
$M$, regarded as divisors in the total space of the canonical line bundle
$X=K_M$. 
The non-compact D-brane worldvolumes in question are the Hirzebruch-Jung
spaces, which are toric minimal resolutions of $(k,l)$ quotient singularities
$\IC^2/\mC_{k,l}$ with the cyclic group $\mC_{k,l}\cong\IZ_k$
parametrized by a pair of coprime integers $(k,l)$ with $k>l>0$. The
intersection matrix $C_{ab}$ in this case is parametrized by a finite
sequence $\{e_i\}_{i=1}^\ell$ of integers $e_i\geq2$, which
appear in the continued fraction expansion
\beq \nonumber
\frac kl =  [e_1,\dots,e_\ell]:= e_1 - \frac{1}{e_2 - \displaystyle{\frac{1}{e_3-\displaystyle{\frac{1}{\ddots \,
    \displaystyle{e_{\ell -1}-\frac1{e_{\ell}}}}}}}}
\eeq
as
\beq \nonumber
C= \big(C_{ab}\big) = \begin{pmatrix} e_1 & -1 & 0 & \cdots &0\\
-1 & e_2  & -1& \cdots &0\\
0 & -1 & e_3 &\cdots&0\\
\vdots &\vdots & \vdots &\ddots&\vdots\\0&0&0&\cdots&e_\ell
\end{pmatrix} \ .
\eeq
We will refer to $C$ as a generalized Cartan matrix. 
The topological data of $M$ are then $\chi(M)=\ell+1$, $b_2^+(M)=\ell$ and
$\Gamma = H^2(M,\IZ)=\IZ^\ell$. Let us construct two sequences of integers
$\{k_i\}_{i=0}^\ell$ and $\{l_i\}_{i=0}^\ell$ defined by the continued fraction expansions
\beq \nonumber
\frac{k_i}{l_i} = [e_1,\dots,e_{i-1}]
\eeq
for $i=2,\dots,\ell+1$, with the initial values
\beq \nonumber
k_0=0 \ , \ k_1=1 \qquad \mbox{and} \qquad l_0=-1 \ , \ l_1=0 \ .
\eeq
In particular, $k_\ell=k$ and $l_\ell=l$. They satisfy the recursion relations
\beq \nonumber
k_{i+1}=e_i\, k_i-k_{i-1} \qquad \mbox{and} \qquad l_{i+1}= e_i\, l_i-l_{i-1}
\eeq
for $i=1,\dots,\ell$.

Let $(\varepsilon_1,\varepsilon_2)$ be
infinitesimal parameters of the abelian subgroup $T=U(1)\times U(1)$ of
the Lorentz group $SO(4)\cong SU(2)\times SU(2)$ of $\IR^4\cong \IC^2$. The action of this torus on
the singular variety $\IC^2/\mC_{k,l}$ lifts to a torus action
on the resolved variety whose weights
$\big(\varepsilon^{(i)}_1,\varepsilon^{(i)}_2)$ on each of the
$\ell+1$ torus-invariant patches
$U_i\cong\IC^2$ of $M$ are given by~\cite{Bonelli12}
\beq \nonumber
\varepsilon^{(i)}_1 = -(k\, l_i-l\, k_i)\, \varepsilon_1-k_i \,
\varepsilon_2 \qquad \mbox{and} \qquad
\varepsilon^{(i)}_2 = (k\, l_{i+1}-l\, k_{i+1})\,
\varepsilon_1+k_{i+1}\, \varepsilon_2
\eeq
for $i=0,1,\dots,\ell$. The $T$-action further lifts to a torus action
on the D-brane moduli spaces (\ref{MnuD}). In the following we will mostly work in the
antidiagonal limit $\varepsilon_1=-\varepsilon_2=: \hbar$ of this
toric action; in this case the holomorphic two-form of $\IC^2$ is
preserved by the $T$-action.

The black hole partition function (\ref{BHZD}) in this instance now
follows from a localization calculation of the equivariant Euler
character $\chi_T({\mathcal M}_{n,u}(M))$ of the moduli space of BPS
configurations in the sector labelled by $(n,u)$;
equivariant localization is used here to properly define characters of the
non-compact instanton moduli space (\ref{MnuD}), and it is understood as a calculation in the
$\Omega$-deformation of the underlying supersymmetric gauge theory wherein the
gauge symmetry is coupled to the abelian isometries generated by
$T$. Localization on the D-brane moduli space (\ref{MnuD}) computes
the partition function (\ref{ZDtauz}) as a sum over the isolated
$T$-fixed points of the Hilbert scheme $M^{[n]}$, which decompose into
contributions from each toric open
set $U_i\cong\IC^2$. Thus the equivariant version of (\ref{ZDtauz})
can be computed from the blow-up formula~\cite[Cor.~5.12]{Gasparim08}
\beq \label{ZDblowup}
\cZ_M(\varepsilon_1,\varepsilon_2; \tau,z) = \sum_{u\in\IZ^\ell}\, q^{\frac12\, u^a\, C_{ab}\, u^b}
\ \prod_{a=1}^\ell\, w_a^{u^a} \ \prod_{i=0}^\ell\,
\cZ_{\IC^2}\big(\varepsilon_1^{(i)},\varepsilon_2^{(i)};\tau\big) \ .
\eeq
The $T$-fixed points of $M^{[n]}$ over $U_i\cong\IC^2$ are parametrized by Young diagrams $Y$ with $|Y|=n$
boxes~\cite{Nakajima99}. The classes which enter in the denominator of the localization
formula in this case cancel against the equivariant Euler characteristic classes at
each fixed point. Hence the contributions from each patch $U_i$ are
independent of the equivariant parameters and coincide with the
generating function for Young diagrams given by Euler's formula
\beq \label{Eulerid}
\cZ_{\IC^2}(\varepsilon_1,\varepsilon_2;\tau) = \sum_{{Y}}\, 
q^{|{{Y}}|} = \hat\eta(\tau)^{-1} \ ,
\eeq
independently of $(\varepsilon_1,\varepsilon_2)$, and the result
(\ref{BHZD}) follows.

In the following we will extend
this reasoning to the computation of quantum black hole degeneracies
which are given by Fourier coefficients of elliptic genera of the
D-brane moduli spaces. Working in an equivariant setting has the
advantage that the computation of the equivariant elliptic genera for
symmetric products, and hence for the D-brane moduli spaces, can be
naturally reduced to infinite product expansions which may be
compared with those from Sect.~\ref{se:40genera}. Let us illustrate
this procedure explicitly for the equivariant version of the refined black hole partition function
(\ref{BHZDyproduct}) which is derived by applying the localization
theorem in $T$-equivariant cohomology to the integral
(\ref{chiycharclass}) over the D-brane moduli
space $X= {\mathcal M}_{n,u}(M)$; it generally depends on the equivariant parameters
$(\varepsilon_1,\varepsilon_2)$ and is given by suitable blow-up
formulas analogous to (\ref{ZDblowup}).
The building block for the blow-up formulas is the partition function on
$M=\IC^2$. By the localization theorem, we obtain the result~\cite{Gasparim08,Hollowood,Li-Liu04}
\beq \label{chiyC2sum}
\cZ_{\IC^2}(t_1,t_2;\tau,\rho) = \sum_{{Y}}\,
q^{|{{Y}}|} \ \prod_{\Box\in Y}\, \left[\frac{\big(1-y\, t_1^{-\ell(\Box)}\, t_2^{a(\Box)+1} 
\big)\, \big( 1-y\, t_1^{\ell(\Box)+1}\, t_2^{-a(\Box)}\big)}{\big(1-t_1^{-\ell(\Box)} \, t_2^{a(\Box)+1} \big)\, \big( 1- t_1^{\ell(\Box)+1} \, t_2^{-a(\Box)}\big)}\right]\,,
\eeq
where
\beq \nonumber
t_1=\e^{-\varepsilon_1} \qquad \mbox{and} \qquad
t_2=\e^{-\varepsilon_2} \ ,
\eeq
while $\ell(\Box)$ and $a(\Box)$ are respectively the leg and arm lengths
of the box $\Box$ of the Young diagram $Y$.
For $y=1$ ($\rho=0$), this agrees with the expected result
(\ref{Eulerid}). In the general case, we can sum the series (\ref{chiyC2sum}) explicitly by
using standard calculations on symmetric product orbifolds. 

\medskip

{\bf Complex plane.} \ 
The Hilbert-Chow morphism $\pi:M^{[n]}\to \mS^n M$ realises the
Hilbert scheme $M^{[n]}$ of $M=\IC^2$ as a semi-small smooth resolution of the
$n$-th symmetric product $\mS^nM$; the torus $T$ acts
diagonally on $\mS^nM$. This identifies the equivariant
$\chi_y$-genera~\cite{Li-Liu04}
\beq 
\chi_y\big((\IC^2)^{[n]} \big)(t_1,t_2) =
\chi_y\big(\frS^n\IC^2\big)(t_1,t_2) \ ,
\eeq
where the $\chi_y$-genus on the right-hand side is the equivariant
orbifold $\chi_y$-genus defined in~\cite[Sect.~3]{Li-Liu04}. The
localization formula applied to (\ref{chiycharclass}) with $X=\IC^2$ yields
\beq \nonumber
\chi_y(\IC^2)(t_1,t_2) = \frac{(1-y\, t_1)\, (1-y\, t_2)}{(1-t_1)\, (1-t_2)}
\ ,
\eeq
and therefore~\cite[Sect.~5.2]{Li-Liu04}
\begin{eqnarray}
\cZ_{\IC^2}(t_1,t_2;\tau,\rho) & = & \exp\Big(\,
\sum_{n=1}^\infty\, \frac{q^n}n\,
\frac{\chi_{y^n}(\IC^2)(t_1^n,t_2^n)}{1-y^n\, q^n}\, \Big)
\nonumber \\[4pt]
& = & 
\prod_{n=1}^\infty \ \prod_{m_1,m_2=0}^\infty \
\left[\frac{\big(1-y^n\, q^n\, t_1^{m_1+1}\, t_2^{m_2}\big)\, \big(1-y^n\,
q^n\, t_1^{m_1}\, t_2^{m_2+1}\big)}{\big(1-y^{n-1}\, q^n\,
t_1^{m_1}\, t_2^{m_2}\big)\, \big(1-y^{n+1}\,  q^n\, t_1^{m_1+1}\,
t_2^{m_2+1}\big)}\right] \nonumber \\[4pt]
& = & 
\!\!\!\!
\prod_{m_1,m_2=0}^\infty \ \left[\frac{\cR\big({\widehat s} = (2\pi
    \ii(\tau + \rho) + \xi_{1,0}^{m_1,m_2})\,
    (1-\ii\varrho(\tau))-2\pi\ii\rho \big)}{\cR\big({\widehat s} = (2\pi \ii\tau+\xi^{m_1,m_2}_{0,0})\,
    (1-\ii\varrho(\tau))-2\pi\ii\rho \big)}
\right]
\label{ZC2t1t2spec} 
\nonumber \\ [4pt]
& \times &
\left[\frac{\cR\big({\widehat s} = (2\pi \ii(\tau + \rho)
  +\xi^{m_1,m_2}_{0,1})\, (1-\ii\varrho(\tau))-
2\pi\ii\rho\big)}
{\cR\big({\widehat s} = (2\pi \ii(\tau + 2\rho) +\xi^{m_1,m_2}_{1,1})\, (1-\ii\varrho(\tau))-
2\pi\ii\rho \big)}\right] \, ,  
\end{eqnarray}
where $\xi^{m_1,m_2}_{a,b} = -(m_1+a)\,
\varepsilon_1-(m_2+b)\, \varepsilon_2$.
Hence in the antidiagonal limit $t_1=t_2^{-1}=t=\e^{-\hbar}$, the
blowup formula (\ref{ZDblowup}) yields
\beq 
\cZ_M(\hbar;\tau,z,\rho) = \Theta_\Gamma (\tau,z) \ \prod_{i=0}^\ell\,
\cZ_{\IC^2}\big(t^{k\, l_i-(l+1)\, k_i}, t^{(l+1)\, k_{i+1}-k\,
  l_{i+1}}\, ;\, \tau,\rho \big) \ .
\eeq
This matches Eq.~(\ref{BHZDyproduct}) in the non-equivariant limit
$t=1$ upon dropping the products over $m_1,m_2$ in
(\ref{ZC2t1t2spec}). Note that in this case, since the geometry is
toric, only the Hodge numbers $h^{j,j}(M)$ for $j=0,1,2$ are
non-vanishing; in particular, all BPS D4-D2-D0 states have vanishing
R-charge $J_0=0$.

\medskip

{\bf Line bundles on $\P^1$.} \ 
Let $M$ be the total space
of the holomorphic line bundle $\cO_{\IP^1}(-k)\to \IP^1$. In this
case $l=1$, $\ell=1$ and $C=e_1=k$, so that the black hole partition function
becomes
\beq 
\cZ_{\cO_{\IP^1}(-k)}(\tau,z)=\frac{\vartheta_3\big(\mbox{$\frac\tau k$},\mbox{$\frac
    zk$}\big)}{\hat\eta(\tau)^2}\,,
\eeq
where $\vartheta_3(\tau,z)$ is the Jacobi elliptic function
\begin{eqnarray}
\vartheta_3(\tau,z) = \prod_{n=1}^\infty\, \big(1-q^n\big)\, \big(1+q^{n-1/2}\,
w \big)\, \big(1+q^{n-1/2}\, w^{-1}\big) \ .
\end{eqnarray}
The spectral function representation is given by combining
(\ref{characterspec}) with
\begin{eqnarray}
\vartheta_3(\tau,z) & = &
\cR\big({\widehat s} = 2\pi \ii\tau(1- \ii\varrho(\tau))\big) \
\cR\big({\widehat s} = \pi\ii(2z - \tau)\, (1-\ii\varrho(\tau)) - 
\ii{\widehat \varrho}(\tau)\big)
\nonumber \\
& \times & \cR\big({\widehat s} = -\pi\ii(2z + \tau)\, (1-\ii\varrho(\tau))
- \ii{\widehat \varrho}(\tau) \big) \ ,
\label{thetaspec}\end{eqnarray}
where ${\widehat \varrho}(\tau) = \pi\, (\varrho(\tau) + \ii)$.
The tangent weights are given by
\beq \nonumber
\varepsilon_1^{(0)}=k \, \varepsilon_1 \ , \ \varepsilon_2^{(0)}=
\varepsilon_2-\varepsilon_1 \qquad \mbox{and} \qquad
\varepsilon_1^{(1)}= \varepsilon_1-\varepsilon_2 \ , \
\varepsilon_2^{(1)} = k \, \varepsilon_2 \ ,
\eeq
and the refined black hole partition function is
\beq \label{ZrefO-k}
\cZ_{\cO_{\IP^1}(-k)}(\hbar;\tau,z,\rho)=\vartheta_3\big(\mbox{$\frac\tau k$},\mbox{$\frac
    zk$}\big) \ \cZ_{\IC^2}\big(t^k,t^{-2}\,;\, \tau,\rho\big)\,
  \cZ_{\IC^2}\big(\, \overline{t}\,^k,\overline{t}\,^{-2}\,;\, \tau,\rho\big)\,,
\eeq
where $\overline{t}=t^{-1}$; its spectral function representation
follows from (\ref{ZC2t1t2spec}) and (\ref{thetaspec}).

\medskip

{\bf ALE spaces.} \ 
The minimal toric Calabi-Yau
resolution of $\IC^2/\IZ_k$ is the ALE space of type $A_{k-1}$. In
this case $l=k-1$,
$\ell=k$ and $e_i=2$ for all $i=1,\dots,k$, so that the intersection
matrix coincides with minus the Cartan matrix of the $A_{k-1}$ Dynkin
diagram. The black hole partition function $\cZ_{A_{k-1}}(\tau,z)$ is then the
character of the
two-dimensional conformal field theory based on the affine Lie algebra
$\widehat\mh\oplus \widehat\frsl(k)_1$~\cite{Vafa}; the level one Kac-Moody
algebra $\widehat\frsl(k)_1$ is derived from the Heisenberg algebra
$\widehat\mh(\Gamma)$ via the Frenkel-Kac construction. The fact that
the generalised chiral algebra is a Kac-Moody algebra in this case is
a consequence of the McKay correspondence.
The tangent weights are given by
\beq \nonumber
\varepsilon_1^{(i)}=(k-i)\, \varepsilon_1-i\, \varepsilon_2 \qquad
\mbox{and} \qquad \varepsilon_2^{(i)}= (-k+i+1)\,
\varepsilon_1+(i+1)\, \varepsilon_2
\eeq
for $i=0,1,\dots,k-1$, and the refined black hole partition function
is
\begin{eqnarray}
\cZ_{A_{k-1}}(\hbar;\tau,z,\rho) &=& \Theta_\Gamma(\tau,z) \ 
\prod_{n,m=0}^\infty \ \left[\frac{\big(1-q^{n+1}\, y^{n+1}\, t^{k\, m}
    \big) \,
\big(1-q^{n+1}\, y^{n+1}\, t^{k\, (m+2)}\big)}
{\big(1-q^{n+1}\, y^n\, t^{k\, (m+1)}\big)\, \big(1-q^{n+1}\,
  y^{n+2}\, t^{k\,(m+1)}\big)}\right]^{k\, (m+1)} \nonumber \\[4pt]
& = & \Theta_\Gamma(\tau,z) \ 
\prod_{m=0}^\infty\, \left[ 
\frac{\cR\big({\widehat s} = (2\pi \ii(\tau + \rho) + \xi^m_0)\,
  (1-\ii\varrho(\tau))- 2\pi\ii\rho \big)}
{\cR\big({\widehat s} = (2\pi \ii\tau + \xi^m_1)\,
  (1-\ii\varrho(\tau))- 2\pi\ii\rho \big)}\right.
\nonumber \\
& \times &
\left.
\frac{\cR\big({\widehat s} = (2\pi \ii(\tau + \rho) + \xi^m_2)\,
  (1-\ii\varrho(\tau))- 2\pi\ii\rho \big)}
{\cR\big({\widehat s} = (2\pi \ii(\tau + 2\rho) + \xi^m_1)\,
  (1-\ii\varrho(\tau))- 2\pi\ii\rho \big)}
\right]^{{k\,(m+1)}}\!\!\!\!\!\!,
\end{eqnarray}
where $\xi^m_{a} = -k\, (m+a)\, \hbar$.
This is the spin character of the two-dimensional conformal
field theory based on the affine Lie algebra $\widehat\mh\oplus
\widehat\frsl(k)_1$; for $k=2$ it coincides with the partition
function (\ref{ZrefO-k}).

\subsection{(2,2) field theory elliptic genus}
\label{2,2}

We now consider microstates of quantum black holes that are enumerated
by elliptic genera of two-dimensional superconformal field theories with
$(2,2)$ worldsheet supersymmetry. 
Underlying all of our black hole partition functions is the elliptic
genus of the D-brane moduli space ${\mathcal M} = {\mathcal M}_{n,u}(M)$ which counts
supersymmetric bound states of D4-D2-D0 systems on a
generic Calabi-Yau threefold. It is given by
the partition function in the Ramond sector of a two-dimensional
$\cN=2$ superconformal sigma-model on the elliptic curve $\cE$ with target manifold
${\mathcal M}$. The equivariant elliptic genus is then defined as a trace
over the Hilbert space $\cH_{n,u}(M)$ in the
Ramond sector as
\beq 
\chi\big({\mathcal M}_{n,u}(M)\,;\, y,p\big)(t_1,t_2) =
\Tr_{\cH_{n,u}(M)}\, (-1)^{F_L} \, y^{J_0}\, p^{L_0}\, t_1^{K_1}\,
t_2^{K_2} \ ,
\eeq
where $(K_1,K_2)$ are the generators of the abelian subgroup
$T=U(1)\times U(1)\subset SU(2)_L\times SU(2)_R$ of the Lorentz group
of $\IR^4$; the extra fugacities $t_1,t_2$ are inserted to further
resolve degeneracies of states. 
The corresponding partition function is then given by the generating
function for elliptic genera
\beq \label{ZDelliptic}
\cZ_M(t_1,t_2;\tau,z,\rho,\sigma) = \sum_{n=0}^\infty \ \sum_{u\in \Gamma} \,
  \chi\big({\mathcal M}_{n,u}(M)\,;\,y,p\big)(t_1,t_2) \ q^n \ \prod_{a=1}^{b^+_2(M)}\,
  w_a^{u^a} \ ,
\eeq
where $p=:\e^{2\pi\ii\sigma}$.
In the non-equivariant case $t_1=t_2=1$, the elliptic genus is the
supersymmetric index given
in terms of multiplicative characteristic classes associated to formal
power series via Eq. (\ref{RRH}) with $X={\mathcal M}_{n,u}(M)$,
$d=2(n-n_u)$ and $q=p$; once again we can explicitly perform the sum
over instanton charges in (\ref{ZDelliptic}) in this instance using
results from Sect.~\ref{HH} and Eq.~(\ref{Ellproduct}) to get
\beq
\cZ_M(\tau,z,\rho,\sigma) = \Theta_\Gamma(\tau,z) \ \prod_{m \geq0
  \,,\,l} \ \prod_{n=1}^\infty \,
\big(1-y^l\, p^m\, q^n\big)^{-\kappa(m\, n,l)} \ ,
\label{ZDellproduct}\eeq
where $\chi(M;p,y)=\sum_{m \geq0,l}\, \kappa(m,l)\,
p^m \, y^l$ is the elliptic genus of $M$.
The equivariant version of (\ref{ZDellproduct}) follows by applying the localization
theorem to the integration in (\ref{RRH}). For $p=0$ it reduces to the Hirzebruch $\chi_y$-genus via (\ref{chiysum}).

To apply the blow-up formulas in this case, we first compute the partition function
(\ref{ZDelliptic}) on $M=\IC^2$ using the localization formula to get~\cite{Gasparim08,Hollowood,Li-Liu04}
\begin{eqnarray}
\cZ_{\IC^2}(t_1,t_2;\tau,\rho,\sigma) & = & \sum_{{Y}}\, \big(y^{-1}\, q
\big)^{|{{Y}}|} 
\nonumber \\ 
& \times & \prod_{\Box\in Y} \ \prod_{m=1}^\infty\, \left[
\frac{\big(1-y\, p^{m-1}\, t_1^{-\ell(\Box)}\, t_2^{a(\Box)+1} \big)\, 
\big(1-y^{-1}\, p^m\, t_1^{\ell(\Box)}\, t_2^{-a(\Box)-1}\big)}
{\big(1-p^{m-1}\, t_1^{-\ell(\Box)}\, t_2^{a(\Box)+1}\big)\, \big(1-
p^m\, t_1^{\ell(\Box)}\, t_2^{-a(\Box)-1}\big)} \right.
\nonumber \\
& \times & \left.
\frac{\big(1-y\, p^{m-1}\, t_1^{\ell(\Box)+1}\, t_2^{-a(\Box)}\big)\,
  \big(1-y^{-1}\, p^m\, t_1^{-\ell(\Box)-1}\, t_2^{a(\Box)}\big)}
{\big(1-p^{m-1}\, t_1^{\ell(\Box)+1}\, t_2^{-a(\Box)}\big)\, \big(1-
  p^m\, t_1^{-\ell(\Box)-1}\, t_2^{a(\Box)}\big)} \right]\,.
\end{eqnarray}
Equivalently, by applying the localization formula to the spectral
function representation (\ref{HRR}) we have
\begin{eqnarray}
\cZ_{\IC^2}(t_1,t_2;\tau,\rho,\sigma) &=& \sum_{{Y}}\, \big(y^{-1}\, q
\big)^{|{{Y}}|} \ \prod_{\Box\in Y} \,
\left[ \frac{\cR\big({\widehat s} =(2\pi\ii\rho +
    \xi^{0,0}_{-\ell(\Box),a(\Box)+1})\, (1-\ii\varrho(\sigma)) \big)
}
{\cR\big({\widehat s} = \xi^{0,0}_{-\ell(\Box),a(\Box)+1}\,
  (1-\ii\varrho(\sigma)) \big)
}\right] \nonumber \\ 
& \times & \left[
\frac{\cR\big({\widehat s} =(2\pi \ii (\sigma-\rho) +
  \xi^{0,0}_{\ell(\Box),-a(\Box) -1})\,
  (1-\ii\varrho(\sigma))\big)}{\cR \big({\widehat s} =(2\pi \ii \sigma
  + \xi^{0,0}_{\ell(\Box),-a(\Box)-1})\, (1-\ii\varrho(\sigma))\big)}
\right] 
\nonumber \\
& \times &
\left[\frac{\cR\big({\widehat s} =(2\pi\ii\rho +
    \xi^{0,0}_{\ell(\Box)+1,-a(\Box)})\, (1-\ii\varrho(\sigma))\big)
}
{\cR\big({\widehat s} =
  \xi^{0,0}_{\ell(\Box)+1,-a(\Box)}\, (1-\ii\varrho(\tau)) \big)
}
\right] 
\nonumber \\ 
& \times &
\left[\frac{\cR\big({\widehat s} =(2\pi \ii (\sigma-\rho) +
    \xi^{0,0}_{-\ell(\Box)-1,a(\Box)})\, (1-\ii\varrho(\sigma))
    \big)}{\cR \big({\widehat s} =(2\pi \ii \sigma +
    \xi^{0,0}_{-\ell(\Box)-1,a(\Box)})\, (1-\ii\varrho(\sigma))\big)}
\right]\,.
\end{eqnarray}
Following~\cite{Dijkgraaf97,Borisov05,Li-Liu04}, we can sum this series
explicitly by identifying this partition
function with the generating function for orbifold
equivariant elliptic genera of symmetric products given by
\beq \nonumber
\cZ_{\IC^2}(t_1,t_2;\tau,\rho,\sigma)  = \sum_{n=0}^\infty \,
\chi\big(\frS^n\IC^2\,;\, y,p\big)(t_1,t_2) \ q^n \ .
\eeq
The localization formula applied to (\ref{RRH}) with $X=\IC^2$ and
$q=p$ yields
\begin{eqnarray}
\chi\big(\IC^2\,;\, y,p\big)(t_1,t_2) &=& 
y^{-1}\, \frac{(1-y\, t_1)\, (1-y\, t_2)}{(1-t_1)\, (1-t_2)} 
\nonumber \\ 
& \times & 
\prod_{m=1}^\infty\, \left[\frac{\big(1-y\, p^{m}\, t_1\big)\, \big(1-y^{-1}\, p^m\,
  t_1^{-1}\big)\,\big(1-y\, p^{m}\, t_2\big)\, \big(1-y^{-1}\, p^m\,
  t_2^{-1}\big)}{\big(1-p^{m}\, t_1\big)\, \big(1-p^m\,
  t_1^{-1}\big)\,\big(1-p^{m}\, t_2\big)\, \big(1-p^m\,
  t_2^{-1}\big)}\right] 
\nonumber \\[4pt]
&=&
y^{-1}\, \frac{(1-y\, t_1)\, (1-y\, t_2)}{(1-t_1)\, (1-t_2)} \ \left[
  \frac{\cR\big({\widehat s} =(2\pi \ii (\sigma+\rho) - \varepsilon_1)\,
    (1-\ii\varrho(\sigma))\big)}{\cR\big({\widehat s} =(2\pi \ii
    \sigma -\varepsilon_1)\, (1-\ii\varrho(\sigma))\big)}\right]
\nonumber \\
& \times & 
\left[\frac{
\cR\big({\widehat s} =(2\pi \ii (\sigma -\rho) +\varepsilon_1)\,
(1-\ii\varrho(\sigma)) \big)}
{
\cR\big({\widehat s} =(2\pi \ii \sigma +\varepsilon_1)\,
(1-\ii\varrho(\sigma))\big) }\right]
\nonumber \\
& \times & 
\left[\frac{ \cR\big({\widehat s} =(2\pi \ii (\sigma+\rho) -
    \varepsilon_2)\, (1-\ii\varrho(\sigma))\big)}{\cR\big({\widehat s}
    =(2\pi \ii \sigma -\varepsilon_2)\,
    (1-\ii\varrho(\sigma))\big)}\right] 
\nonumber \\ 
& \times & 
\left[\frac{
\cR\big({\widehat s} =(2\pi \ii (\sigma -\rho) +\varepsilon_2)\, (1-\ii\varrho(\sigma))\big)}
{
\cR\big({\widehat s} =(2\pi \ii \sigma +\varepsilon_2)\,
(1-\ii\varrho(\sigma))\big)} \right]\,.
\end{eqnarray}
By~\cite[Thm.~3.1]{Li-Liu04} one then has
\beq 
\cZ_{\IC^2}(t_1,t_2;\tau,\rho,\sigma)  = \exp\Big(\,
\sum_{n,n'=1}^\infty\, \frac{q^{n\,n'}}{n'}\ \frac1n \, \sum_{j=0}^{n-1}\,
  \chi\big(\IC^2\,;\, y^{n'},\e^{2\pi\ii j\, n'/n}\, p^{n'/n}\big)(t^{n'}_1,t^{n'}_2)\,
  \Big) \ .
\eeq
Let us work out the corresponding free energy. For this, we
define integers $\kappa(l,m,k_1,k_2)\in\IZ$ by the expansion
\beq 
&& \prod_{m=1}^\infty\,\left[
\frac{\big(1-y\, p^{m}\, t_1\big)\, \big(1-y^{-1}\, p^m\,
  t_1^{-1}\big)\,\big(1-y\, p^{m}\, t_2\big)\, \big(1-y^{-1}\, p^m \,
  t_2^{-1}\big)}{\big(1-p^{m}\, t_1\big)\, \big(1-p^m \,
  t_1^{-1}\big)\,\big(1-p^{m}\, t_2\big)\, \big(1-p^m \,
  t_2^{-1}\big)}\right] 
\nonumber \\ 
= \!\!\!\!\!\!\! &&  
\sum_{l=0}^\infty \ \sum_{m,k_1,k_2\in\IZ} \, \kappa(l,m,k_1,k_2) \ p^l\,
y^m\, t_1^{k_1}\, t_2^{k_2}\,.
\label{cnmkexp}\eeq
Since the left-hand side of (\ref{cnmkexp}) is invariant under the changes of variables
$t_1\leftrightarrow t_2$ and $y\to y^{-1}, t_1\to t_1^{-1},t_2\to
t_2^{-1}$, these integers have the symmetry properties
\beq \nonumber
\kappa(l,m,k_1,k_2) =\kappa(l,m,k_2,k_1)= \kappa(l,-m,-k_1,-k_2) \ .
\eeq
We thus have
\begin{eqnarray}
\log \cZ_{\IC^2}(t_1,t_2;\tau,\rho,\sigma) & = &
\sum_{n,n'=1}^\infty\, \frac{\big(y^{-1} \, q^n \big)^{n'}}{n'}\,
\frac{\big(1-y^{n'}\, t_1^{n'}\big)\, \big(1-y^{n'}\,
  t_2^{n'}\big)}{\big(1-t_1^{n'}\big)\, \big(1-t_2^{n'}\big)} 
\nonumber \\
& \times & 
\frac1n \, \sum_{j=0}^{n-1} \ \sum_{l=0}^\infty \ 
\sum_{m,k_1,k_2\in\IZ}\, \kappa(l,m,k_1,k_2)\, \e^{2\pi\ii l\, j/n} \,
p^{l\, n'/n}\, y^{m\, n'}\, t_1^{k_1\, n'}\, t_2^{k_2\, n'} 
\nonumber \\[4pt]
& = & 
\sum_{n,n'=1}^\infty\, \frac{\big(y^{-1}\,
q^n\big)^{n'}}{n'}\, \big(1-y^{n'}\, t_1^{n'}-y^{n'}\,
t_2^{n'}+y^{2n'}\, t_1^{n'}\, t_2^{n'} \big)  \ 
\sum_{r_1,r_2=0}^\infty\,  t_1^{n'\, r_1}\, t_2^{n'\, r_2}
\nonumber \\
& \times &
\sum_{l=0}^\infty \ 
\sum_{m,k_1,k_2\in\IZ}\, \kappa(n\, l,m,k_1,k_2) \, 
p^{l\,n'}\, y^{m\,n'}\, t_1^{k_1\, n'}\, t_2^{k_2\, n'}
\nonumber \\[4pt]
& = &
-\sum_{l,r_1,r_2=0}^\infty \ \sum_{m,k_1,k_2\in\IZ} \ 
\sum_{n=1}^\infty \, \kappa(n\,l,m,k_1,k_2) 
\nonumber \\
& \times & \log \left[\left(\frac{1-q^n\, y^{m-1}\,
    p^l \,
    t_1^{k_1+r_1}\, t_2^{k_2+r_2}}{1-q^n\, y^m\, p^l \,
    t_1^{k_1+r_1+1}\, t_2^{k_2+r_2}}\right)\right. 
\nonumber \\
& \times & \left. 
\left(\frac{
1-q^n\, y^{m+1}\, p^l\, t_1^{k_1+r_1+1}\, t_2^{k_2+r_2+1} }
{1-q^n\, y^m\, p^l \, t_1^{k_1+r_1}\, t_2^{k_2+r_2+1}}\right)\right] 
\end{eqnarray}
and we finally get
\begin{eqnarray}
\cZ_{\IC^2}(t_1,t_2;\tau,\rho,\sigma) &=&
\prod_{\stackrel{\scriptstyle l, r_1, r_2 \geq0\,,\,n>0}{
\scriptstyle m, k_1, k_2\in{\mathbb Z}}} \ 
\left[\left(\frac{1-y^m\, p^l\, q^n \, t_1^{k_1+r_1+1} \,
t_2^{k_2+r_2}}{1-y^{m-1}\, p^l \, q^n \, t_1^{k_1+r_1} \,
t_2^{k_2+r_2}}\right) \right. 
\nonumber \\ 
& \times & \left. 
\left(\frac{1-y^m\, p^l \, q^n \, t_1^{k_1+r_1}\, 
t_2^{k_2+r_2+1}}{1-y^{m+1}\, p^l\, q^n\, t_1^{k_1+r_1+1} \,
t_2^{k_2+r_2+1}}\right) \right]^{\kappa(n\,l,m,k_1,k_2)}\,.
\label{final1}
\end{eqnarray}

The blow-up formula (\ref{ZDblowup}) in the antidiagonal limit yields
\beq 
\cZ_M(\hbar;\tau,z,\rho,\sigma) = \Theta_\Gamma(\tau,z) \ \prod_{i=0}^\ell\,
\cZ_{\IC^2}\big(t^{k\, l_i-(l+1)\, k_i}, t^{(l+1)\, k_{i+1}-k\,
  l_{i+1}}\, ;\, \tau,\rho,\sigma \big) \ .
\eeq
For example, when $M$ is the total space of the holomorphic line
bundle $\cO_{\IP^1}(-k)\to \IP^1$, we get
\beq 
\cZ_{\cO_{\IP^1}(-k)}(\hbar;\tau,z,\rho,\sigma)=\vartheta_3\big(\mbox{$\frac\tau k$},\mbox{$\frac
    zk$}\big) \ \cZ_{\IC^2}\big(t^k,t^{-2}\,;\, \tau,\rho, \sigma \big)\,
  \cZ_{\IC^2}\big(\, \overline{t}\,^k,\overline{t}\,^{-2}\,;\,
  \tau,\rho,\sigma\big) \ .
\eeq
On the other hand, when $M$ is the ALE space of type $A_{k-1}$
we find
\begin{eqnarray}
\cZ_{A_{k-1}}(\hbar;\tau,z,\rho,\sigma) &=& \Theta_\Gamma(\tau,z) \ 
\prod_{\stackrel{\scriptstyle l,r\geq0\,,\,n>0 }{
\scriptstyle m, j\in{\mathbb Z}}} \ 
\left[ \left(\frac{1-y^m\, p^l\, q^n\,
    t^{k\, (j+1) }}{1-y^{m-1} \, p^l\, q^n\, t^{k\,j } }\right) \right. 
\nonumber \\
& \times &
\left. \left(\frac{1-y^m\, p^l \, q^n\, t^{k\,(j-1) }}{
    1-y^{m+1} \, p^l\, q^n\, t^{k\, j}}\right) \right]^{k\,
  (r+1)\, \kappa(n\,l,m,j-r)}\,,
\label{final2}
\end{eqnarray}
where $\kappa(n,m,j):=\kappa(n,m,k_1-k_2,0)$ in the expansion (\ref{cnmkexp})
with $t_1=t_2^{-1}=t$.

The elliptic genera {\rm (\ref{final1})} and {\rm (\ref{final2})} can be
described in terms of a set of Kerov's symmetric functions {\rm
  \cite{Kerov}}. Using the Cauchy identity for Kerov's functions in
the Hall-Littlewood case and Ruelle-type spectral functions, one can
obtain the corresponding spectral partition functions in their final
form analogously to our M-theory partition functions. Similar
calculations exist in the literature for the matrix elements of the currents of the quantum affine 
algebra $U_q(\widehat{\frsl}(2))$ (see e.g. \cite{King}). The
derivations and resulting formulas are somewhat cumbersome to describe
and lie outside the scope of the present paper; we plan to address this problem elsewhere.

As we show in Appendix~\ref{Pole}, the elliptic genus can be continued to
an entire function without essential singularities. It would be interesting to understand its relation
to the elliptic genus of the $(4,0)$ superconformal field theory in
the M-theory frame. This presumably entails constructing partition
functions which depend on moduli which measure fluctuations of the
D4-brane around the divisor class of its worldvolume $M$ in $X$, and similarly for
the D2-branes; in this region of the moduli space one must study the
full nonlinear Dirac-Born-Infeld theory as the worldvolume effective
theory, and the connected
components of the D-brane moduli space are expected to be singular
fibrations over the Hilbert schemes $M^{[n]}$. Then one should find a sequence of duality transformations to a particular chamber
of the moduli space where the D4-brane
decays into a D6 brane-antibrane pair, with the D6-brane
worldvolume theory governed by a suitable refinement or
categorification of
Donaldson-Thomas theory (or equivalently topological string theory) as
in~\cite{Aganagic12}; via this chain of dualities the $AdS_2\times
S^2\times X$ near horizon geometry is related to Type~IIA string
theory on $X$ with a D6 brane-antibrane pair, where the M-theory lift of the D6-brane is a Taub-NUT
geometry.

Note that such walls of marginal stability can in principal
arise by computing degeneracies from Fourier coefficients of the
elliptic genus as in Eq.~(\ref{Fouriercontour}); the moduli dependence is due to the moduli dependence
of the Fourier integration contours and the pole structure of the
partition functions described in Appendix~\ref{Pole}. Moving around the
moduli space corresponds to deforming the contour and crossing a wall
in the moduli space corresponds to crossing a pole of the partition
function. The jump in the degeneracy upon crossing a wall from one
domain to another is given by the residue at the pole which is
crossed, as computed in Appendix~\ref{Pole}. It would be interesting
to see if the equivariant deformation of the elliptic genera computed
here could be used in this way to reproduce our M-theory partition
functions by analysing the pole structures in the deformation
parameters $(t_1,t_2)$.

\section{Conclusions}

In this paper we discussed various applications of affine Lie algebra 
representations, Hilbert schemes, elliptic genera and their
generalizations to the computation of M-theory and D-brane quantum
partition functions for microscopic black hole ensembles. A 
central concept in all examples considered in this paper is that of 
elliptic genera. The notion of elliptic genus was introduced in 
\cite{Ochanine} with applications to quantum field theory in \cite{Witten87}. It has been argued 
that it is possible to use elliptic modular forms to write generating 
functions of quantum field theory as infinite series of operators 
associated with homologies of finite-dimensional Lie algebras. The 
elliptic genus can be interpreted as a natural invariant in a generalized 
cohomology theory, called elliptic cohomology 
\cite{Landweber,Landweber95}, which can be regarded as a 
extension of K-theory. From this point of view these partition functions might 
be related to elliptic cohomology and K-theory.
All examples considered in this paper point towards this conjectural
link. We have shown that 
elliptic genera, in the context of black hole partition functions,
can be converted into products of spectral functions associated with 
$q$-series which inherit the (co)homology properties of appropriate 
(poly)graded Lie algebras.
This common feature, that quantum 
generating functions can be reproduced in terms of spectral functions, encodes the connection with infinite-dimensional Lie 
algebras and their homologies, together with the remarkable link to hyperbolic geometry.

\subsection*{Acknowledgments}

AAB would like to acknowledge the Conselho Nacional
de Desenvolvimento Cient\'ifico e Tecnol\'ogico (CNPq, Brazil) and Funda\c cao Araucaria 
(Parana, Brazil) for financial support. The support of the Academy of Finland under the 
Projects No. 136539 and 140886 is gratefully acknowledged. 
The work of RJS was partially supported by the Consolidated Grant
ST/J000310/1 from the UK Science and Technology Facilities Council,
and by Grant RPG-404 from the Leverhulme Trust.

\appendix

\section{Spectral functions of hyperbolic three-manifolds \label{Spectral}}

The Euclidean sector of $AdS_3$ has a quotient description as a closed oriented
hyperbolic three-manifold $\IH^3/\mG^\gamma$. The
complex unimodular group $G=SL(2, {\mathbb C})$ acts on the real
hyperbolic three-space $\IH^3= \{(x,y,z) \in\IR^3 \ | \ z>0\}$ 
in the standard way: for $(x,y,z)\in \IH^3$ and 
$$
g=\begin{pmatrix} a& b\\c &d \end{pmatrix} \ \in \ G\,,
$$
one has
$g\triangleright (x,y,z)= (u,v,w)\in
\IH^3$, where
\beq
u+\ii v =
\frac{(a\,r+b)\, \overline{(c\, r+d)}+ a\, \overline{c}\, z^2}{|c\, r+d|^2 +
|c|^2\, z^2} \qquad \mbox{and} \qquad
 w = \frac z{|c\, r+d|^2 + |c|^2\, z^2} \ , \nonumber
\eeq
with $r=x+\ii y$. Let $\mG^\gamma\subset G$ be
the infinite discrete Schottky subgroup of $G$ defined as
\begin{eqnarray}
\mG^\gamma = \big\{{\gamma}^n \ \big| \ n\in {\mathbb Z} \big\}\ , \qquad
\mbox{with} \quad {\gamma} = {\rm diag}\big(\e^{2\pi
  \ii\overline{\tau}} \,,\, \e^{-2\pi\ii\overline{\tau}} \, \big)\,,
\label{group}
\end{eqnarray}
for $\tau\in\IC^+$; this subgroup
acts on $\IH^3$ as a rotation through angle $2\pi\,{\rm Re}\, \tau$ in
the $(x,y)$-plane and a dilatation by $\e^{2\pi\,{\rm Im}\, \tau}$.

One can construct a Selberg-type zeta-function for any subgroup 
$\mG^\gamma := \mG_{(a, b)}^\gamma$ generated by a single hyperbolic
element of the form $\gamma:={\gamma_{(a, b)}} = {\rm diag}(\e^z, \e^{-z})$,
where $z= a+\ii b$ with $a, b >0$; for the case of $AdS_3$ we specialise to $a = 2\pi\,
{\rm Im}\,\tau$ and $b = 2\pi \, {\rm Re}\,\tau$. Then the
Patterson-Selberg spectral function $Z_{\mG^\gamma} (s)$ which can be
attached to the hyperbolic three-manifold $\IH^3/\mG^\gamma$ (with
acyclic orthogonal representation of $\pi_1(\IH^3/\mG^\gamma)$) has the form
\begin{equation}
Z_{\mG^\gamma}(s) :=\prod_{k_1,k_2=0}^\infty\, \big(1-\e^{\ii b\,
(k_1-k_2)}\, \e^{-a\,(k_1+k_2+s)} \big) \ .
\label{zeta00}
\end{equation}
The zeroes of $Z_{\mG^\gamma} (s)$ are precisely the set of complex numbers 
$$
\zeta_{n,k_{1},k_{2}} = -\left(k_{1}+k_{2}\right)+\ii\left(k_{1}-
k_{2}\right) \mbox{$\frac ba$} + \mbox{$\frac{2\pi \ii n}a$}\,,
$$
with $n \in {\mathbb Z}$. The magnitude of the zeta-function is bounded
for both ${\rm Re}\,s\geq 0$ and ${\rm Re}\,s\leq 0$, and its growth can be estimated as
\begin{equation} 
\big|Z_{\mG^\gamma}(s) \big| \leq \Big(\,
\prod_{k_1+k_2\leq
|s|}\,  \e^{|s|\, \ell}\, \Big)\,
\Big(\, \prod_{k_1+k_2\geq
|s|}\, \big(1- \e^{(|s|-k_1-k_2)\, \ell} \big)\,\Big)
\leq C_1\,\e^{C_2\, |s|^3}
\label{estimate}
\end{equation}
for suitable constants $\ell,C_1, C_2$. The first product on the right-hand side of (\ref{estimate}) gives the exponential growth, while the second product is bounded. 
The spectral function $Z_{\mG^\gamma} (s)$ is an entire function of
order three and of finite type which can be written as a Hadamard product~\cite{Perry}
\begin{equation}
Z_{\mG^\gamma}(s) = 
\e^{Q(s)} \
\prod_{\zeta \in {\Sigma}}\, 
\Big(\, 1-\frac{s}{\zeta}\, \Big)\, 
\exp \Big(\,
\frac{s}{\zeta} + \frac{s^2}{2\zeta^2} +
\frac{s^3}{3\zeta^3}\, \Big)\ ,
\label{Hadamard}
\end{equation}
where $\Sigma$ is the set of zeroes $\zeta := \zeta_{n,k_{1},k_{2}}$
and $Q(s)$ is a polynomial of degree at most three. From the Hadamard
product representation (\ref{Hadamard}) it follows that
\begin{equation}
\frac{\dd}{\dd s}\, {\rm log}\,Z_{\mG^\gamma} (s) =
\frac{\dd}{\dd s} Q(s) + \sum_{\zeta \in \Sigma}\,
\frac{(s/\zeta)^3}{s-\zeta}\ .
\end{equation}
Let us define $\Xi(y \pm \ii\xi):= \frac\dd{\dd s} \, {\rm log}\, Z_{\mG^\gamma} (s)$
for $s = y \pm \ii\xi$. Then
\begin{equation}
\Xi (y \pm \ii\xi)
= \frac{\dd}{\dd s}Q(s= y \pm \ii\xi)
- \ii \, \sum_{y \pm \ii\varepsilon \in \Sigma}\
\frac{(y \pm \ii\xi)^3}
{(y \pm \ii\varepsilon)^3\, (\pm \, \xi- \varepsilon)}\ .
\label{Phi}
\end{equation}

Let us introduce next the Ruelle zeta-functions $\cR(s)$
\cite{Bytsenko07,Bo-By,BBE}. The function ${\mathcal R}(s)$ is a ratio
of Patterson-Selberg zeta-functions given by $\cR(s) :=
Z_{\mG^\gamma}(s)\, Z_{\mG^\gamma}(s+2)/Z_{\mG^\gamma}(s+1)$; it is defined for ${\rm Re}\,
s\gg1$ and can be
continued to a meromorphic function on the entire complex plane $\mathbb
C$. Its value $\cR(0)$ computes the $L^2$-analytic torsion of the hyperbolic
three-manifold $\IH^3/\mG^\gamma$, and in the case of $AdS_3$ one has the infinite product identities:
\begin{eqnarray}
\prod_{n=m}^{\infty}\, \big(1- q^{\mu\, n+\varepsilon} \big) 
& = & \frac{Z_{\mG^\gamma}(s)}{ Z_{\mG^\gamma}\big(s+ \mu\,(1 +
  \ii\varrho(\tau))\big)} \nonumber \\[4pt] &=& \cR\big(s = (\mu \,m + \varepsilon)\,
(1-\ii\varrho(\tau)) + 1-\mu\big) \ ,
\\[4pt]
\prod_{n=m}^{\infty}\, \big(1+ q^{\mu\,n+\varepsilon} \big)
& = & \frac{Z_{\mG^\gamma}(s)}{ Z_{\mG^\gamma}\big(s+ \mu\,(1 +
  \ii\varrho(\tau))\big)} \nonumber\\[4pt] &= & \cR\big(s = (\mu \,m + \varepsilon)\,
(1-\ii\varrho(\tau)) + 1-\mu+ \ii/(2\,{\rm Im}\,\tau)\big)\ , 
\end{eqnarray}
where $q:= \e^{2\pi \ii\tau}$, $\varrho(\tau) = {\rm Re}\,\tau/{\rm Im}\,\tau$, 
$\mu\in\IR$, $m\geq1$ and $\varepsilon \in {\mathbb C}$. For $\nu\in
{\mathbb C}$, we can use the Ruelle functions $\cR(s)$ to naturally write more
general infinite product identities:
\begin{eqnarray}
\prod_{n=m}^{\infty}\,\big(1-q^{\mu\,n+ \varepsilon}\big)^{\nu\,n} & = & 
\cR\big(s=(\mu\, m + \varepsilon)\, (1-\ii\varrho(\tau))+1-\mu\big)^{\nu\, m}
\nonumber \\
&\times& \prod_{n=m+1}^{\infty} \,
\cR\big(s=(\mu\, n + \varepsilon)\, (1-\ii\varrho(\tau))+1-\mu
\big)^{\nu}\ ,
\label{RU1}
\\[4pt]
\prod_{n=m}^{\infty}\, \big(1+q^{\mu\,n+ \varepsilon}\big)^{\nu\, n} & = & 
\cR\big(s = (\mu \,m + \varepsilon)\,
(1-\ii\varrho(\tau)) + 1-\mu+ \ii/(2\,{\rm Im}\,\tau)\big)^{\nu\,m}
\label{RU2}\\
&\times& \prod_{n=m+1}^{\infty} \,
\cR\big(s = (\mu \,n + \varepsilon)\,
(1-\ii\varrho(\tau)) + 1-\mu+ \ii/(2\,{\rm Im}\,\tau)\big)^{\nu}\ .
\nonumber 
\end{eqnarray}

\section{Singularities and poles of elliptic genera}
\label{Pole}

As a function of $y$, the elliptic genus $\chi(M;q,y)$ in
(\ref{Chern1}) has infinitely many simple poles at the values $y = q^{1-m+v}$ and $y =
q^{m+v}$,  where $v= c_1([M])/2\pi \ii\tau$ and $m$ is a positive
integer. One can calculate the residues of these simple poles. For
this, let us set
\begin{equation}
\cP(y) := \prod_{n=1}^\infty \, \frac{\big(1-q^{n-1-v}\big)\,
  \big(1-q^{n+v}\big)}{\big(1-y\, q^{n-1-v}\big)\, \big(1- y^{-1}\,
  q^{n+v} \big)}\ .
\end{equation}
Then
\begin{eqnarray}
{\rm Res} \big(\cP(y)\,,\, q^{1-m+v}\big) & = & \lim_{ y\rightarrow
  q^{1-m+v}}\,  
\big(y-q^{1-m+v}\big) \, \cP(y) 
\\[4pt]
& = & -q^{1-m+v}\ \prod_{n\neq m} \ 
\frac{1}{1-q^{n-m}} \ \prod_{n=1}^\infty \,
\frac{\big(1-q^{n-1-v}\big)\, \big(1-q^{n+v}\big)}{1-q^{n+m-1}}\ . \nonumber
\end{eqnarray}
Note that
\begin{equation}
\prod_{n\neq m} \ \frac{1}{1-q^{n-m}}
= \prod_{n=1}^{m-1}\, \frac{1}{1-q^{n-m}} \ \prod_{n = m+1}^\infty \ \frac{1}{1-q^{n-m}}
= \frac1{\big(q^{-m}; q\big)_{m-1}\ \cR\big(s = 1-\ii\varrho(\tau) \big)}
\end{equation}
where $(a; q)_m := \prod_{j=1}^m\, (1-a\, q^j)$, and therefore
\begin{eqnarray}
{\rm Res} \big(\cP(y)\,,\, q^{1-m+v}\big) & = &  -\, \frac{q^{1-m+ v}}{\big(q^{-m}; q\big)_{m-1}}
\\
&\times& \left[\frac{\cR\big(s=- v\, (1-\ii\varrho(\tau))\big)\
    \cR\big({s}=(1+v)\, (1-\ii\varrho(\tau))\big)}
{\cR\big({s}= 1-\ii\varrho(\tau)\big)\ \cR\big(s= m\,
  (1-\ii\varrho(\tau)) \big)}\right] \ . \nonumber
\end{eqnarray}
An easy way of calculating the residue at
$y= q^{m+v}$ is to replace $v\rightarrow
v-1$, $m\rightarrow -m$ to get
\begin{eqnarray}
{\rm Res} \big(\cP(y)\,,\, q^{m+v}\big) & = &  -\, \frac{q^{m+ v}}{\big(q^{m}; q\big)_{m-1}}
\\
&\times& \left[\frac{\cR\big({s}=- v\,(1-\ii\varrho(\tau))\big)\
    \cR\big({s}=(1+ v)\, (1-\ii\varrho(\tau))\big)}
{\cR\big({s}= (2m+1)\, (1-\ii\varrho(\tau))\big)\ \cR\big({s}= m\,
  (1-\ii\varrho(\tau)) \big)}\right] \ . \nonumber
\end{eqnarray}
Hence we can construct an entire function which has no essential
singularities in the complex plane (except possibly at
infinity). Similar pole structures are contained in the explicit
formulas for all elliptic
genera of Sects.~\ref{EllGenera}, \ref{SUGRAelliptic}, \ref{Refined} and~\ref{2,2}. In
these examples one can use a subtraction procedure to remove
singularities, some elements of which have been presented here.

\end{document}